\documentclass[onecolumn,amsmath,amssymb,11pt,superscriptaddress,nofootinbib]{revtex4}
\newlength{\oldoddsidemargin}
\oldoddsidemargin=\oddsidemargin \oddsidemargin=\evensidemargin
\evensidemargin=\oldoddsidemargin

\makeatletter
\def\cleardoublepage{\clearpage\if@twoside \ifodd\c@page\else
   \hbox{}
   \thispagestyle{empty}
   \newpage
   \if@twocolumn\hbox{}\newpage\fi\fi\fi}
\makeatother \clearpage{\pagestyle{plain}\cleardoublepage}

\usepackage{tabularx}
\usepackage{longtable}

\usepackage{multirow}
\usepackage{graphicx, color}
\usepackage{subfigure}
\usepackage{amsmath}
\usepackage{amsfonts}
\usepackage{amssymb}
\usepackage[english,german]{babel}
\usepackage{bbding}

\usepackage{float}
\usepackage{dsfont}   % unit matrix symbol (\mathds{1}), or real numbers \mathds{R} ...
\usepackage{hyperref}
\usepackage{mathrsfs}
\usepackage{makeidx}
\usepackage{listings}
\makeindex

\usepackage{setspace}
\newcolumntype{C}[1]{>{\centering\arraybackslash}b{#1}}
\newcolumntype{D}[1]{>{\centering\arraybackslash}b{#1}}
\newcolumntype{R}[1]{>{\raggedleft\arraybackslash}b{#1}}

\newcolumntype{M}[1]{>{\centering\arraybackslash}m{#1}}

\def\e{\mathrm{e}}

\def\ln{\mathrm{ln}}
\def\p{\partial}
\def\d{\mathrm d}

\def\bea{\begin{eqnarray}}
\def\eea{\end{eqnarray}}

\def\I{\mathrm{i}}

\def\bsm{\left( \!\begin{smallmatrix}}
\def\esm{\end{smallmatrix} \!\right)}

\newcommand{\mac}[1]{\mathcal{#1}}

\newcommand{\nn}{\nonumber}

\def\cE{\mac E}
\def\cT{\mac T}

\numberwithin{equation}{section}

\renewcommand{\a}{\alpha}

\newcommand{\Th}{\Theta}

\newcommand{\R}{\mathbb{R}}

\newcommand{\be}{\begin{equation}}
\newcommand{\ee}{\end{equation}}
\newcommand{\ba}{\begin{align}}
\newcommand{\ea}{\end{align}}

\def\cR{\mac{R}}
\def\cD{\mac{D}}

\def\bcD{\bar{\mac{D}}}
\def\Phid{\Phi^\dagger}

\def\cDla{{\mac{D}}_{{\alpha}}}

\def\cDlbP{{\mac{D}}_\beta\Phi}

\def\cDua{{\mac{D}}^{{\alpha}}}

\def\cDubP{{\mac{D}}^\beta\Phi}

\def\bcDla{\bar{\mac{D}}_{\dot{\alpha}}}

\def\bcDua{\bar{\mac{D}}^{\dot{\alpha}}}

\def\bcDlb{\bar{\mac{D}}_{\dot{\beta}}}

\def\bcDubPd{\bar{\mac{D}}^{\dot{\beta}}\Phid}

\def\Sib{\cD^\beta\Phi\cD_\beta\Phi\bcD_{\dot\beta}\Phid\bcD^{\dot\beta}\Phid}

\def\weyl{\stackrel{\text{\tiny WEYL}}{\longrightarrow}}

\def\Phone{G}
\def\phone{\mac{G}}

\def\zf{{\text{0f}}}

\def\blc{\big|}
\def\Blc{\Big|}

\graphicspath{{gfx/}}

\begin{document}
\allowdisplaybreaks
\selectlanguage{english}
\begin{titlepage}

\title{A Cosmological Super--Bounce}

\author{Michael Koehn}
\email[]{michael.koehn@aei.mpg.de}
\author{Jean-Luc Lehners}
\email[]{jlehners@aei.mpg.de}
\affiliation{Max-Planck-Institute for Gravitational Physics (Albert-Einstein-Institute), 14476 Potsdam, Germany}
\author{Burt A. Ovrut}
\email[]{ovrut@elcapitan.hep.upenn.edu}
\affiliation{Department of Physics, University of Pennsylvania,\\ 209 South 33rd Street, Philadelphia, PA 19104-6395, U.S.A.}

\begin{abstract}

\vspace{.3in}
\noindent
We study a model for a non-singular cosmic bounce in ${\cal N}=1$ supergravity, based on supergravity versions of the ghost condensate and cubic Galileon scalar field theories. The bounce is preceded by an ekpyrotic contracting phase which prevents the growth of anisotropies in the approach to the bounce, and allows for the generation of scale-invariant density perturbations that carry over into the expanding phase of the universe. We present the conditions required for the bounce to be free of ghost excitations, as well as the tunings that are necessary in order for the model to be in agreement with cosmological observations. All of these conditions can be met. Our model thus provides a proof-of-principle that non-singular bounces are viable in supergravity, despite the fact that during the bounce the null energy condition is violated.
\end{abstract}
\maketitle

\end{titlepage}
\tableofcontents

\section{Introduction}
A fundamental question of modern cosmology is whether our currently expanding universe had a beginning, presumably in the form of an initial classically singular event (perhaps with space and time emerging from a more abstract and fundamental description), or whether the expanding phase was preceded by a phase of contraction (perhaps with phases of expansion and contraction alternating to yield a cyclic cosmology). In the present paper, we wish to study the latter possibility. More specifically, we are interested in models where the reversal from contraction to expansion occurs in an entirely non-singular manner already at the classical level, with the scale factor of the universe smoothly reaching a minimum value well above the Planck length before starting to grow again. 

It is well-known that, in a flat Friedmann-Lema$\hat{\imath}$tre-Robertson-Walker (FLRW) universe, the requirement that the Hubble rate $H$ increase requires the matter constituents of the universe to violate the null energy condition, i.e. the sum of energy density $\rho$ and pressure $p$ must become negative in order for a bounce to occur (general discussion of bounces are provided in e.g. \cite{Novello:2008ra,Lehners:2011kr}). In recent years, effective scalar field theories have been constructed which have the remarkable property that they allow for violations of the null energy condition without however causing any obvious pathologies: these are the ghost condensate \cite{ArkaniHamed:2003uy} and Galileon \cite{Nicolis:2008in,Nicolis:2009qm,Deffayet:2009wt} models\footnote{There also exist non-singular bounce models in theories with infinite numbers of derivatives, see e.g. \cite{Biswas:2005qr,Biswas:2010zk,Biswas:2012bp}.}. However, what has remained largely unclear is whether such models also make 
sense from a more fundamental perspective. For instance, it remains unclear whether these models can be derived from string theory \cite{Donagi:2001fs,Khoury:2012dn,Ovrut:2012wn,Easson:2013bda}. In the present paper, we wish to take a step in this direction, by studying models of a non-singular bounce in ${\cal N}=1$ supergravity. Since supergravity theories enjoy remarkable stability properties, and since they arise as low-energy approximations to string theory, it is certainly of interest to analyze whether these theories allow for non-singular bounces. As we will show, one can indeed embed models of non-singular bounces in supergravity, although a number of fine-tunings are required in order to render the models free of ghosts and observationally viable.

In our bounce model, we include a description of an ekpyrotic phase in supergravity. This is a crucial element in making the model viable. The reason for this is the following.  When all matter components have an equation of state which is such that the pressure is smaller than the energy density, $p<\rho,$ then anisotropies get amplified in a contracting universe with the result that the universe undergoes BKL oscillations and collapses in a chaotic big crunch \cite{Belinsky:1982pk}. In this case, a cosmic bounce cannot occur, as the curvatures build up to such an extent that gravitational collapse is unavoidable. However, in the presence of ekpyrotic matter, which is characterized by an ultra-stiff equation of state $p > \rho,$ anisotropies are suppressed and it becomes meaningful to consider a transition to a non-singular bounce phase \cite{Khoury:2001wf,Khoury:2001bz,Erickson:2003zm} (see \cite{Xue:2013bva} for a recent non-perturbative numerical study of these issues). Thus, if one wants to avoid having 
to assume highly special initial conditions for the contracting phase of the universe, an ekpyrotic phase is required, irrespective of whether the cosmological perturbations are also generated via ekpyrosis. For this reason, we believe that it is crucial to combine the description of a non-singular bounce with an ekpyrotic phase.

Our paper is structured as follows: we will start by presenting our model without supersymmetry in section \ref{sectionmodel}. There, we also include the results of a numerical computation in order to show a representative solution explicitly. We then extend the model to supergravity in section \ref{sectionsugra}, dealing with the bounce and ekpyrotic phases in turn. The perturbations and associated stability properties of the model are analyzed in section \ref{sectionstability}. We present our conclusions in section \ref{sectiondiscussion}. The component expansion of the full supergravity Lagrangian is presented separately in an Appendix -- we note that this also constitutes the first construction of a Galileon Lagrangian in (old-minimal) supergravity.

\section{The cosmological model} \label{sectionmodel}

We will start by describing our model in a non-supersymmetric framework first, before discussing its embedding in supergravity. The model we consider is based on the bounce model developed by Cai et al. \cite{Cai:2012va} and consists of a scalar field $\phi$ with non-canonical kinetic terms and a potential $V(\phi).$ In natural units ($8\pi G = M_{Pl}^{-2}=1$), the Lagrangian is given by
\be
{\cal L} = \sqrt{-g}\big( -\frac{\cR}{2} +P(X,\phi) + g(\phi) X \Box \phi \big), \label{Lagrangian}
\ee
where $\cR$ is the Ricci scalar and
\be
P(X,\phi) = k(\phi) X + \tau(\phi) X^2 - V(\phi)
\ee
with $X \equiv -\frac{1}{2}(\partial\phi)^2.$ The explicit forms of the functions $k,\tau,g,V$ are specified below. The term proportional to $g(\phi)$ is the first non-trivial Galileon Lagrangian. We will take the background to be a flat FLRW universe, with metric
\be
ds^2 = - dt^2 + a(t)^2 \delta_{ij} dx^i dx^j \ .
\ee
The energy density and pressure are then given by
\bea
\rho &=& \frac{1}{2}k(\phi)\dot\phi^2 + \frac{3}{4}\tau(\phi)\dot\phi^4 - 3 g(\phi) H \dot\phi^3 + \frac{1}{2}\dot{g}(\phi)\dot\phi^3 +V(\phi) \, , \\
p &=& \frac{1}{2}k(\phi)\dot\phi^2 + \frac{1}{4}\tau(\phi)\dot\phi^4 + g(\phi) \dot\phi^2 \ddot\phi + \frac{1}{2}\dot{g}(\phi)\dot\phi^3 - V(\phi),
\eea
where we have restricted to time dependence only. The Einstein equations reduce to the Friedmann equations
\bea
3H^2 &=& \rho \, , \\
\dot{H} &=& - \frac{1}{2}(\rho + p) \, ,
\eea
and the scalar equation of motion is given by
\begin{align}
0=&P_{,\phi}-P_{,X}(\ddot\phi+3H\dot\phi)-P_{,XX}\ddot\phi\dot\phi^2-P_{,X\phi}\dot\phi^2\nn\\
&+g(\phi)(6\ddot\phi\dot\phi H+9H^2\dot\phi^2+3\dot H\dot\phi^2)-2g_{,\phi}\dot\phi^2\ddot\phi-\frac{1}{2}g_{,\phi\phi}\dot\phi^4 \ .
\end{align}

The idea of our model is as follows: at large positive values of $\phi,$ the universe starts to undergo an ekpyrotic contraction phase, with approximate potential $V(\phi) \approx - V_0 e^{-c \phi},$
with $c > \sqrt{6}$ so that the equation of state of the scalar field is $w=p/\rho > 1.$ During this phase, the kinetic term is approximately canonical and the universe contracts slowly while anisotropies are suppressed. Around $\phi=\phi_{ek-end},$ the potential bottoms out and rises back up to zero. At that time, the universe goes over into a kinetic phase, i.e. a phase where the energy density is dominated by the kinetic energy of the scalar field and the potential becomes irrelevant. Subsequently, the ordinary kinetic term switches sign while the higher-derivative terms proportional to $X^2$ and $X \Box \phi$ are switched on simultaneously. Both the effective ghost condensate (${\cal L} \sim -X+X^2$) and the Galileon term contribute to a brief violation of the null energy condition (NEC), such that the universe can undergo a bounce at small values of $\phi.$ After the bounce, the universe goes over into a standard expanding phase, while the kinetic term becomes canonical once more. We are assuming that 
reheating takes place around the time of the bounce, and that this causes the universe to become filled with radiation. The ordinary hot big bang cosmological model follows.

\begin{figure}
\includegraphics[width=0.75 \textwidth]{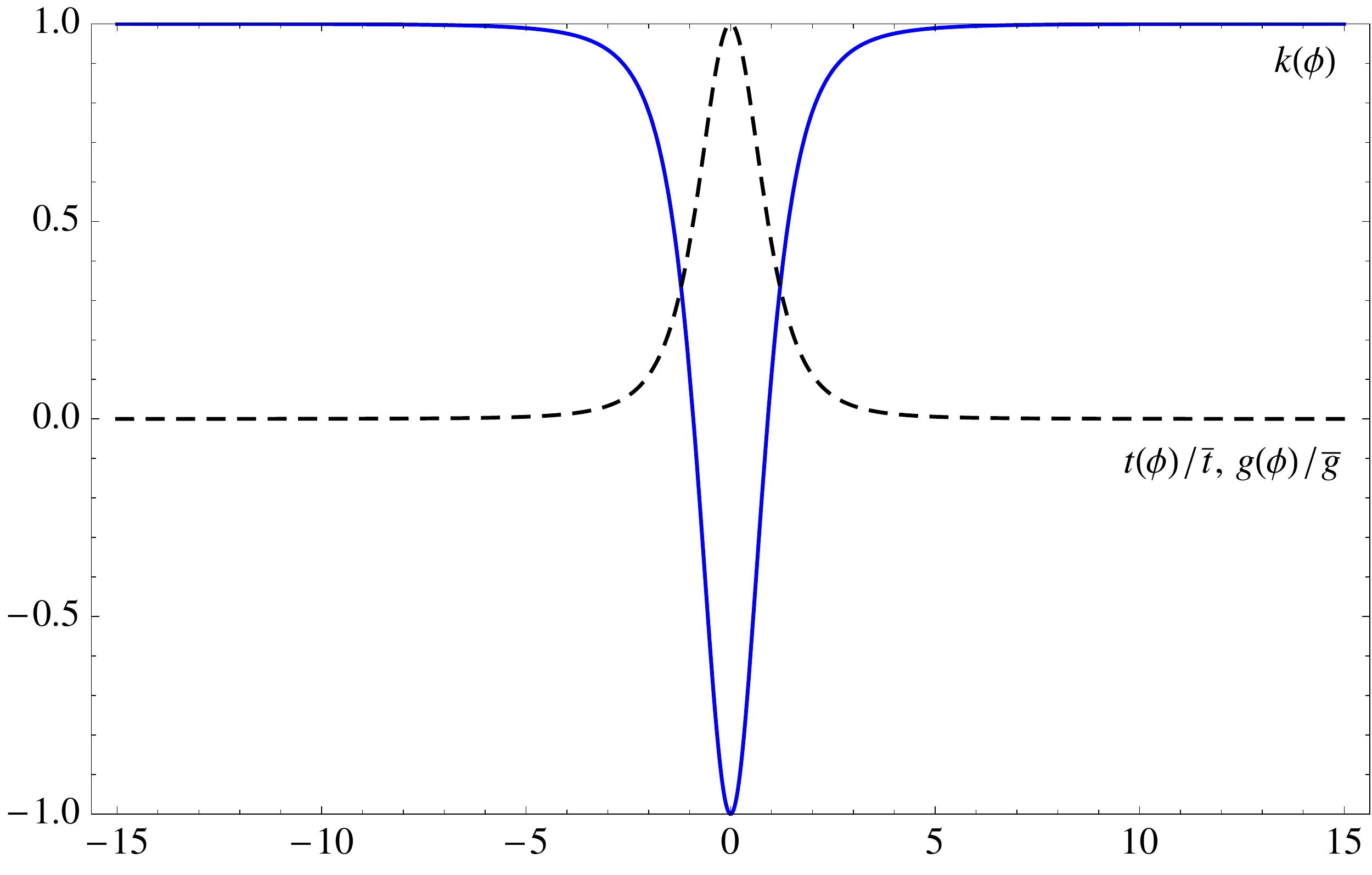}
\caption{The solid curve shows $k(\phi)$ while the dashed curve shows the normalized functions $\tau(\phi)/\bar{\tau},g(\phi)/\bar{g}$, all with $\kappa=1/4.$ }\label{fig:kaat}
\end{figure}

Let us now be a little more specific. We are choosing the kinetic function $k(\phi)$ to be equal to unity everywhere except near $\phi=0$, where it smoothly switches sign and where the bounce occurs. We use the specific form (chosen to allow for a simple supersymmetric extension later on)
\be
k(\phi) = 1 - \frac{2}{(1+2\kappa \phi^2)^2},
\ee
see Fig. \ref{fig:kaat} for an illustration. Here $\kappa$ denotes a parameter that controls the width in field space over which the kinetic term switches sign. The function $\tau(\phi)$ controls the strength of a term that is the square of the ordinary kinetic term, and $g(\phi)$ determines the strength of the Galileon term. We are choosing both such that they interpolate between $0$ and the constants $\bar{t},\bar{g}$ while the ordinary kinetic term switches sign,
\be
\tau(\phi) = \frac{\bar{\tau}}{(1+2\kappa \phi^2)^2}, \quad g(\phi) = \frac{\bar{g}}{(1+2\kappa \phi^2)^2},
\ee
also see Fig. \ref{fig:kaat}. It is crucial that these functions are already non-zero when $k(\phi)$ passes through zero, otherwise a singularity would develop at this point. (Note that, when $k(\phi)$ reaches zero, the higher-derivative term $\tau(\partial \phi)^4$ can act like an ordinary kinetic term $\sim - \tau\dot\phi^2  (\partial \phi)^2$ because the background is non-trivial, $\dot\phi \neq 0.$) In the model presented in \cite{Cai:2012va}, the higher-derivative terms were always ``on'', i.e. the choice $\tau(\phi)=\bar{\tau},g(\phi)=\bar{g}$ was made. This is however not necessary in order to achieve a bounce. In fact, the higher-derivative terms play a completely negligible role during the ekpyrotic phase, but would significantly complicate the supersymmetry analysis during that phase. Hence, we are only turning them on during the bounce phase. We should emphasize that the specific functions written out above are chosen for convenience only -- there is considerable freedom in these choices, and, in 
particular, 
the functional forms of $k,\tau,g$ need not be related to each other in a simple manner like they are in our example.

\begin{figure}
\includegraphics[width=0.75 \textwidth]{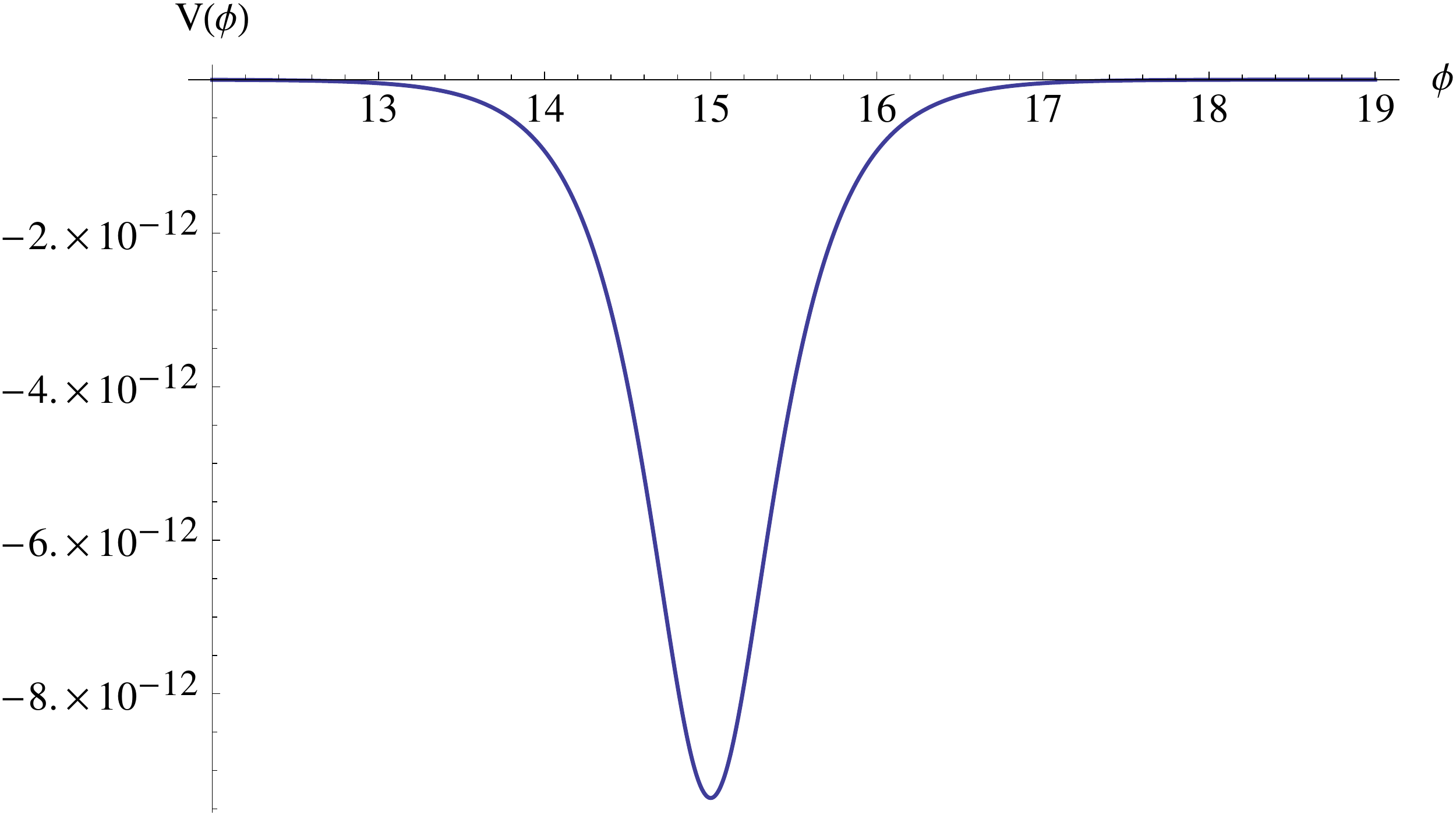}
\caption{The ekpyrotic potential. The ekpyrotic phase starts at large positive $\phi,$ with the field rolling down the potential towards smaller values of the field. Around $\phi_{ek-end}\approx 15$ the potential starts to come back up to zero, and is irrelevant from then on. The bounce occurs at small values, $\phi \approx 0$.}\label{fig:ekpot}
\end{figure}

At large values of $\phi,$ we are choosing an ekpyrotic potential
\be
V(\phi) = - V_0 v(\phi) e^{-c(\phi) \phi}
\ee
where $v(\phi)$ is a function chosen such that the potential turns off for $\phi < \phi_{ek-end}$. One can take, for example, $v(\phi) = \frac{1}{2}[1+\tanh(\lambda(\phi-\phi_{ek-end}))]$ for some positive constant $\lambda$ -- also see Fig. \ref{fig:ekpot}. Here $c(\phi)$ is a slowly varying function of $\phi,$ with $c(\phi)> \sqrt{6}$ over a significant field range. Then the background solution is given by the ekpyrotic scaling solution
\be
a(t)\propto (-t+t_*)^{2/c^2} \qquad
\phi(t)=-\frac{2}{c}\ln \left(-\left({\frac{c^2 V_0}{c^2-6}}\right)^{1/2} (t-t_*)\right), \label{eq:ek}
\ee
for some constant $t_*$ which would correspond to the time of the big crunch if the ekpyrotic phase were to continue until that time. Note that in the solution above time runs from large negative vaues of $t$ towards smaller negative values. During the ekpyrotic phase, the equation of state of the scalar field is given by
\be
w = \frac{p}{\rho} = \frac{\dot\phi^2 - 2V}{\dot\phi^2 + 2V} \approx \frac{c(\phi)^2}{3}-1 > 1, \label{eq:w}
\ee
which implies that the ekpyrotic scalar field energy density (which grows as $a^{-3-3w}$) grows faster than any other component of the total energy density, in particular faster than the growth of the energy stored in the homogeneous curvature ($\propto a^{-2}$) and anisotropic curvature ($\propto a^{-6}$). In this way, the universe becomes flat and smooth, and the approximation of a flat FLRW background is justified.

At $\phi=\phi_{ek-end}= \phi(t_{ek-end}),$ the potential bottoms out and comes back up to zero again. From then on, the potential becomes irrelevant and the ekpyrotic phase goes over into a kinetic phase described by the approximate solution
\be
a(t) \propto (-t+t_0)^{1/3}, \qquad \phi(t) -\phi_0 = -\sqrt{\frac{2}{3}}\ln(-t+t_0),
\ee
where $t_0,\phi_0$ are constants, with $t_0$ representing the time of the would-be big crunch if the higher-derivative terms were absent, while $\phi_0 = \phi_{ek-end}+\sqrt{2/3}\ln(t_0-t_{ek-end})$ is determined by matching onto the ekpyrotic solution above. During the kinetic phase, the equation of state is given by $w=1,$ so that anisotropies remain small while the homogeneous curvature is further suppressed.

\begin{figure}
\includegraphics[width=0.75 \textwidth]{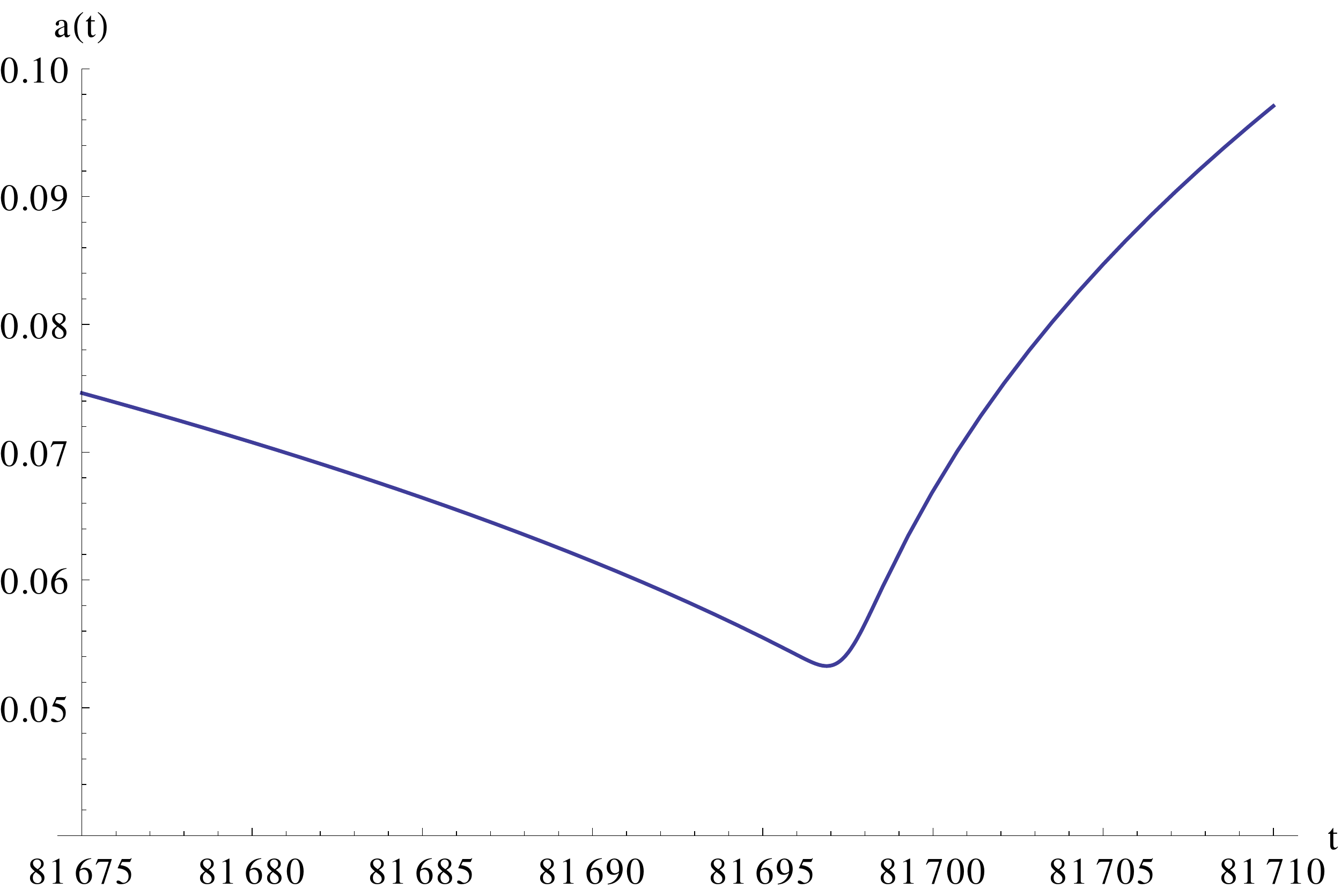}
\caption{The scale factor around the time of the bounce. Our numerical evaluation starts at $\phi_0=17/2$ with $\dot\phi_0=-10^{-5},$ $a_0=1$ and $H_0$ is determined by the Friedmann equation. We are using the parameters $\kappa=1/4,\bar{\tau}=1,\bar{g}=1/100.$ The figure shows a zoom-in on the most interesting time period, namely that of the bounce. One can clearly see that the bounce is smooth. The next three figures plot the evolution of various quantities during that same time period.}\label{fig:ScaleFactor}
\end{figure}

\begin{figure}
\includegraphics[width=0.75 \textwidth]{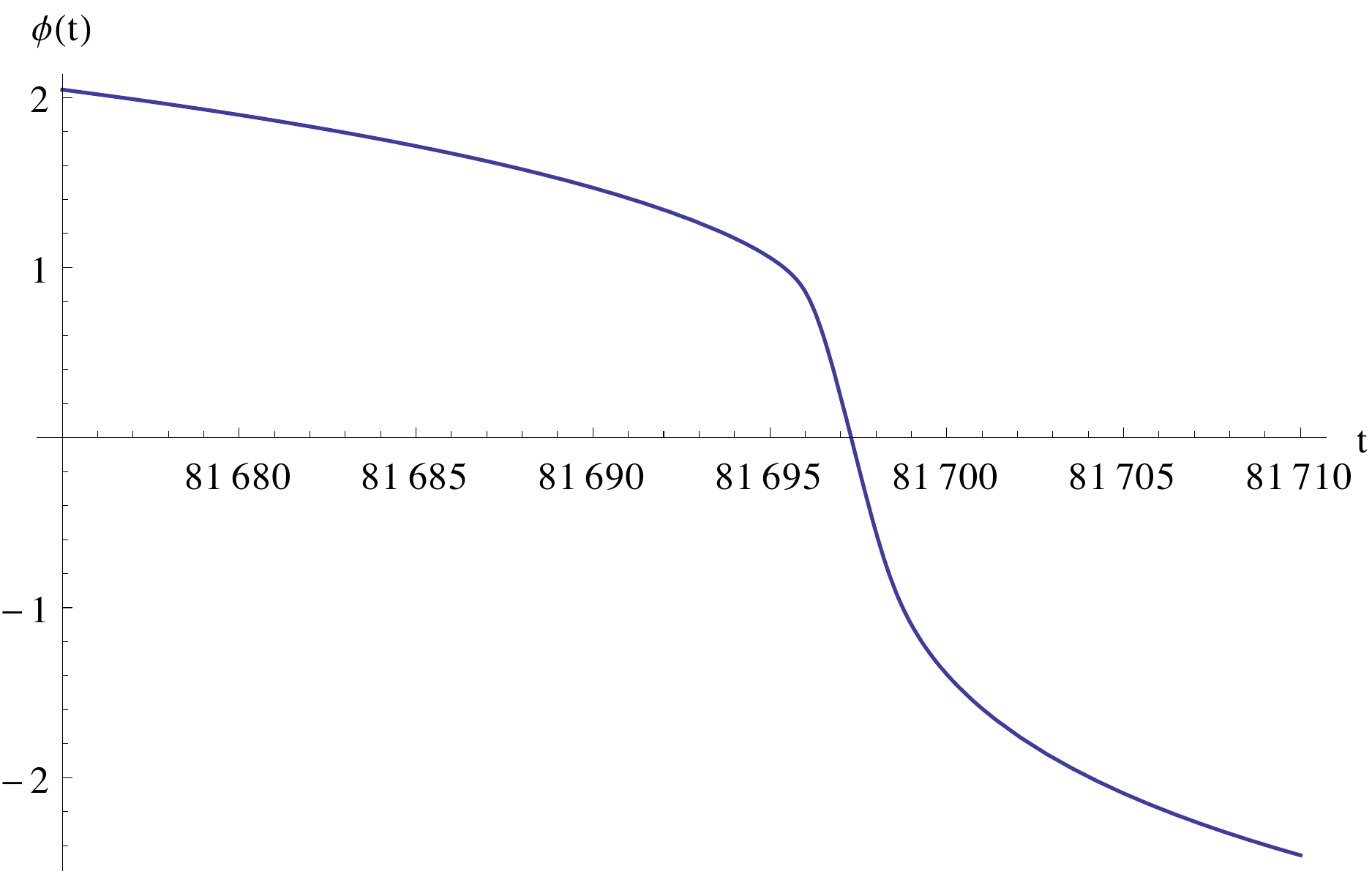}
\caption{The evolution of the scalar field $\phi$ during the bounce phase. The approximately linear evolution near $\phi=0$ corresponds to the ghost condensate phase which is responsible for the bounce.}\label{fig:Phi}
\end{figure}

\begin{figure}
\includegraphics[width=0.75 \textwidth]{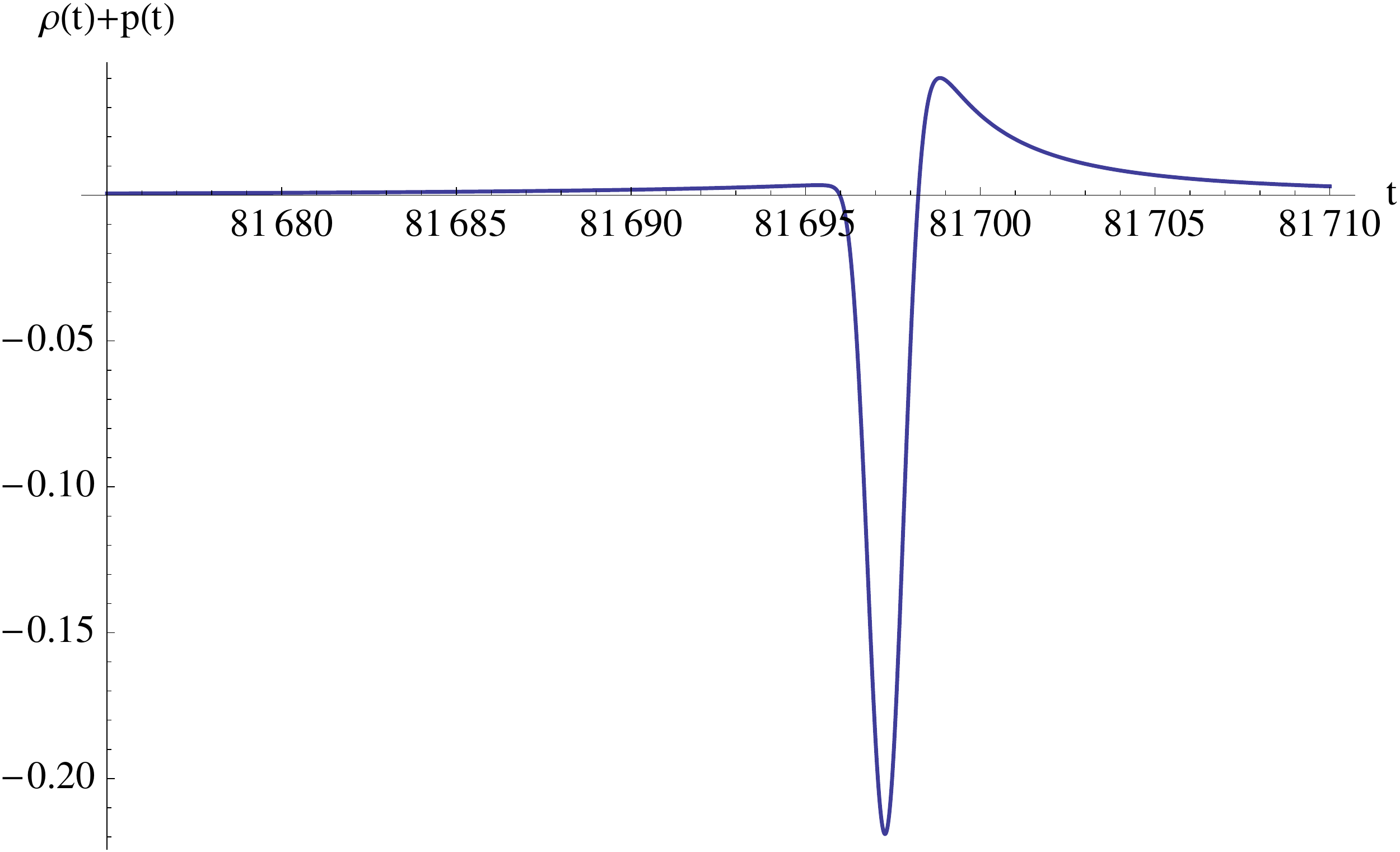}
\caption{The sum of energy density and pressure during the bounce phase. When this quantity goes negative, the null energy condition is violated -- this is a necessary condition for a non-singular bounce in a flat FLRW universe.}\label{fig:NEC}
\end{figure}

\begin{figure}
\includegraphics[width=0.75 \textwidth]{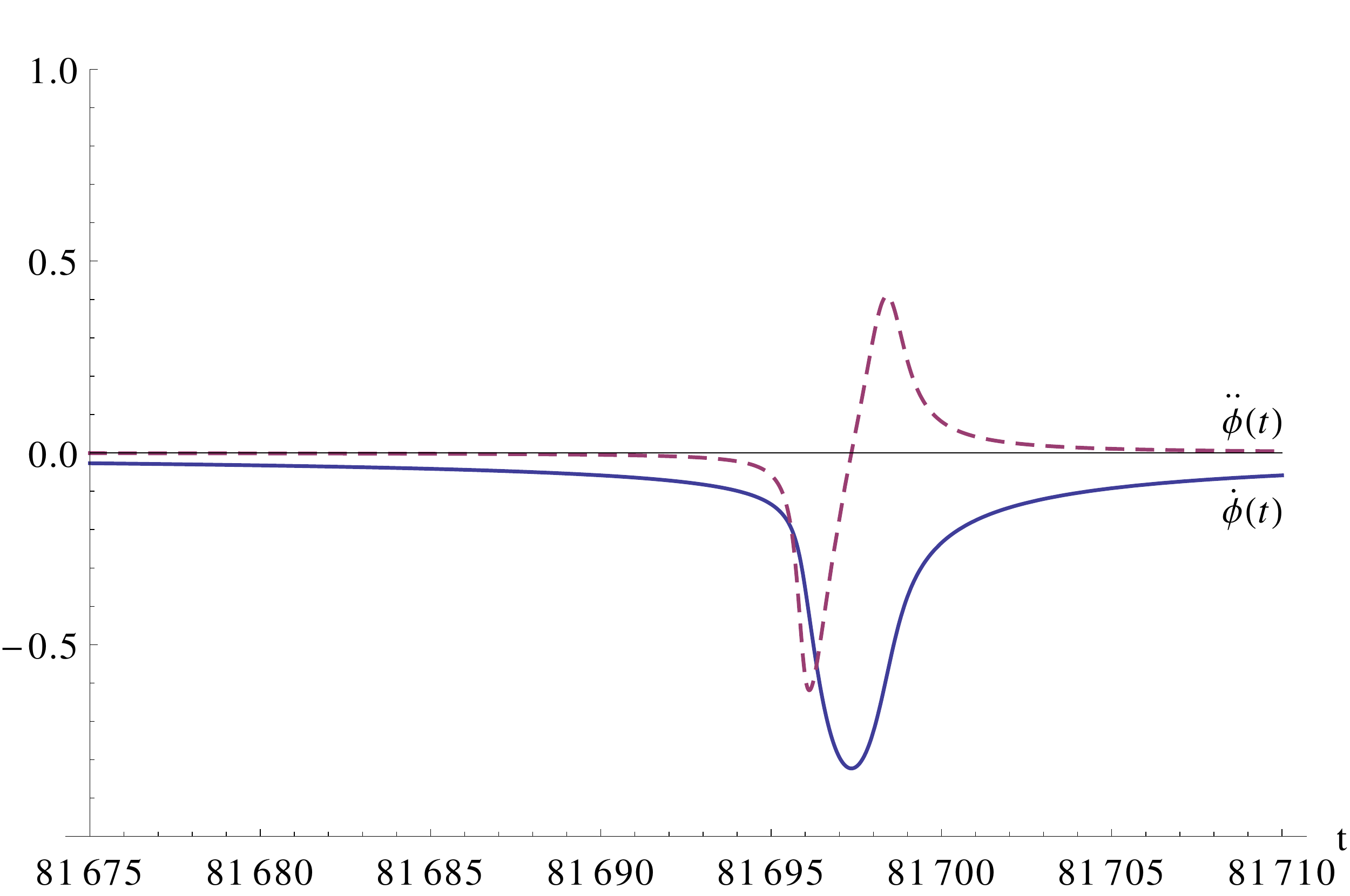}
\caption{The evolution of $\dot\phi$ (solid curve) and $\ddot\phi$ (dashed curve) during the bounce. In our example, the actual value of the terms involving $\ddot\phi$ are very small, since they are multiplied by an additional factor of $\bar{g}=1/100.$ Since the figure demonstrates that the hierarchy $|\ddot\phi| \lesssim |\dot\phi| < 1$ is satisfied throughout, we can see that our effective field theory treatment is fully justified, as explained in the main text. }\label{fig:PhiDot}
\end{figure}

As the scalar is nearing $\phi=0,$ its kinetic term starts switching sign while the higher-derivative terms become important. The Einstein equations imply that $\dot{H}= - \frac{1}{2}(\rho + p),$ i.e. that the Hubble rate can only increase if the sum of energy density and pressure is negative. This is the same condition as that for a violation of the NEC. Only when this sum is negative can the universe revert from contraction to expansion in a non-singular way. Thus a necessary condition for a bounce to occur is that we must have
\be
\rho + p = k(\phi)\dot\phi^2 + \tau(\phi)\dot\phi^4 + g(\phi) \dot\phi^2 \ddot\phi - 3 g(\phi) H \dot\phi^3 + \dot{g}(\phi)\dot\phi^3 < 0
\ee
over a sufficiently long time span. In the present paper, we are pricipally interested in the case where the ghost condensate terms dominate over the Galileon, i.e. we consider the case $\tau(\phi) \gg g(\phi).$ Then, at the moment where $k(\phi)$ reaches $-1,$ the condition for NEC violation translates into
\be
\dot\phi^2 < \frac{1}{\bar{\tau}} \qquad \textrm{for NEC violation at bounce.}
\ee
But this condition needs to be satisfied in any case, since $1/\bar{\tau}$ represents the cut-off of the ghost condensate theory -- if $\dot\phi^2$ were larger than $1/\bar{\tau}$ the $X^2$ term would be larger than the $X$ term and we would lose control over our effective theory. Assuming a short kinetic phase, the field velocity at the onset of the bounce phase is essentially determined by the velocity at the end of the ekpyrotic phase, $\dot\phi_{ek-end}^2 \approx 2|V_{ek-end}|$. In turn this implies that the depth of the ekpyrotic potential must also be below the cut-off scale of the ghost condensate for our model to be viable. Thus we must satisfy the hierarchy
\be
|V_{ek-end}| < \frac{1}{\bar{\tau}} < M_{Pl}^4,
\ee
where $M_{Pl}$ denotes the Planck mass.

Figs. \ref{fig:ScaleFactor} - \ref{fig:PhiDot} present an explicit numerical example of the bounce phase. The numerical evaluation is started after the ekpyrotic phase has come to an end, i.e. at the time when the kinetic phase is underway and about to go over into the bounce phase.  As the figures show, a smooth bounce is obtained during the time period that the NEC is violated. Furthermore, we note that during the time that the NEC is violated, the scalar field evolves almost exactly linearly with time -- this is a characteristic feature of ghost condensation. 

As one can see in Fig. \ref{fig:PhiDot}, the scalar field reaches its largest velocities during the bounce period, and so it is during this period that we must verify that our effective field theory treatment is consistently applicable. Crucially, the cut-off scale of the ghost condensate (set to $1$ in the  numerics) is not surpassed by $\dot{\phi},$ and thus the $X^2$ term is everywhere subdominant to the $X$ term. This is, in fact, due to the presence of the Galileon term, and motivates its inclusion in our model. If the Galileon were absent, then at the moment of the bounce the $X^2$ term would become just as large as the $X$ term, and thus we would come to doubt the validity of our effective field theory precisely at the most crucial moment. Since the Galileon also contributes to violating the NEC, in its presence the $X^2$ term need not become as large. In our numerical example we have chosen the coefficient of the Galileon term to be small, $\bar{g} = \bar{\tau}/100.$ However, the strength of the 
Galileon term could easily be increased, and this would only reinforce the present arguments. The upper limit on $\bar{g}$ is determined by two factors: first, the energy scale $1/\bar{g}^{4/3}$ should not lie below the ghost condensate scale $1/\bar{\tau}.$ And secondly, one must ensure that the scalar field derivatives remain below the regime of validity of the Galileon term itself, i.e. we need $|\dot\phi| < 1/\bar{g}^{2/3}, |\ddot\phi| < 1/\bar{g}$. As the plot of $\ddot{\phi}$ shows, in our example the second time derivative of $\phi$ is smaller than the first derivative, and thus the Galileon term is consistently small throughout. This fact also implies that it is consistent to neglect possible additional higher-order terms (assuming they do not have unnaturally large coefficients), as they would include further factors of the field with various numbers of derivatives, with all of these factors being small. Thus, we conclude that our analysis is trustworthy.

\section{Extension to supergravity} \label{sectionsugra}

Now we would like to extend the model that we have just presented to minimal ${\cal N}=1$ supergravity. By this, we mean that we would like to obtain a supergravitational action which, when all the extra fields required by supersymmetry are set to zero, reduces to the action \eqref{Lagrangian} above. Our construction is based on the results of \cite{Koehn:2012ar}, where we developed the formalism required for coupling chiral superfields with higher-derivative kinetic terms to four-dimensional ${\cal{N}}=1$ supergravity\footnote{Also see \cite{Farakos:2012qu}, where similar results were obtained. Closely related works include \cite{Khoury:2010gb,Khoury:2011da,Buchbinder:1988yu,Buchbinder:1994iw,Banin:2006db,Brandt:1993vd,Brandt:1996au,Antoniadis:2007xc}. For an application to DBI inflation, see \cite{Koehn:2012np}. A review is given in \cite{Lehners:2013cta}.}. In \cite{Koehn:2012te} we applied the formalism to a pure ghost condensate model -- many results from that work will be incorporated below. Since we 
are interested in cosmological applications, we will neglect fermionic component fields throughout. 

The action is formulated in curved superspace, and we are using the conventions of Wess and Bagger \cite{Wess:1992cp}. In the following, we only provide a brief review of the construction of supergravity theories -- for a thorough discussion see \cite{Wess:1992cp}. A chiral superfield $\Phi$ is characterized by the expansion
\be
\Phi = A  + \Th^\a \Th_\a F,
\ee
where $A,F$ are two complex scalar fields, with $F$ typically playing the role of an auxiliary field with non-propagating degrees of freedom. The $\Th$ coordinates are Grassmann-valued and carry local Lorentz indices ($\a$ denotes the index of a two-component Weyl spinor) -- they extend ordinary spacetime to curved superspace. Supersymmetric Lagrangians can be constructed from the chiral integrals
\be
\int \d^2 \Th (\bcD^2 - 8 R) L +H.c.,
\ee
where $L$ is a scalar, hermitian function. The chiral projector in curved superspace is $\bcD^2 - 8R,$ where $\bcD_{\dot\a}$ is a spinorial component of the curved superspace covariant derivative $\cD_A=\{ \cD_a,\cD_\a,\bcD_{\dot\a}\}.$ The curvature superfield $R$ admits the component expansion
\be
R = -\frac16 M + \Th^2 \big( \frac{1}{12}\cR -\frac19 MM^* - \frac{1}{18} b_m b^m + \frac16 \I {e_a}^m \cD_m b^a\big) \ ,
\ee
where $\cR$ is the Ricci scalar. The complex scalar $M$ and the real vector $b_m$ are the auxiliary fields of
supergravity. We will also employ the chiral density $\mac{E}$ with expansion
\be
2 \mac{E} = e (1 - \Th^2 M^*),
\ee
where $e$ is the determinant of the vierbein. One can relate the tangent space Lorentz indices $A=\{ a,\a,\dot\a\}$ to the spacetime indices $M=\{ m,\mu,\dot\mu\}$ via the supervielbein ${E_M}^A$ and its inverse, with ${E_m}^a ={e_m}^a$ being the ordinary vierbein.

Our construction is built on the superspace Lagrangian
\begin{align}
\mac{L}=\int \d^2\Theta2\cE\big[W(\Phi)-\frac18(\bcD^2-8R)\big(&-3\e^{-K(\Phi,\Phid)/3}+\Sib T\nn\\
&+\cD^\beta\Phi\cDlbP\bcDlb\bcDubPd\Phone\big)\big]+H.c.
\label{LagrangeSuperspace}
\end{align}
We will explain the meaning of the functions $W,K,T,\Phone$ in turn. $W(\Phi)$ is the superpotential, and consists of a holomorphic function of $\Phi.$ The
K\"{a}hler potential $K(\Phi,\Phi^{\dagger})$ is a hermitian function which determines the two-derivative kinetic term for the lowest component $A$ of $\Phi.$ In the absence of higher-derivative terms, $W$ and $K$ also determine the potential of the theory. Here, we have two additional terms: the first is proportional to the tensor superfield $T(\Phi,\Phid,\cD_m\Phi,\cD_n\Phid,\ldots),$ which is an arbitrary function of the chiral and anti-chiral superfields and their covariant derivatives, with all indices contracted. The tensor nature of this superfield refers to its transformation properties under field redefinitions, and need not concern us here (see \cite{Koehn:2012ar}, where this term was first introduced, and where it is described in great detail). This term is crucial for obtaining a supergravity extension of the ghost condensate \cite{Koehn:2012te}, and moreover plays a crucial role in obtaining a model that is devoid of perturbative ghost instabilities, as we will see. The final term in the 
Lagrangian is required in order to obtain a supergravity extension of the Galileon term, and contains an arbitrary function $\Phone(\Phi,\Phid)$ of the chiral and anti-chiral superfields. The embedding of the Galileon Lagrangian into supergravity is new to the literature.
 
The full component expansion of the above action is lengthy - we are presenting it in detail in the Appendix. However, our construction is designed specifically such that it simplifies considerably in the two regimes of interest, namely in the ekpyrotic phase where the higher-derivative terms are unimportant but where the potential plays a crucial role, and in the bounce phase where the potential is unimportant but the higher-derivative terms essential. We will discuss these two regimes separately, starting with the bounce phase.

\subsection{The bounce}

During the bounce phase, the superpotential is effectively zero, $W \approx 0,$ and as shown in the Appendix the Lagrangian then reduces to the component form
\begin{align}
\frac1{e}\mac{L}_{bounce}=&-\frac12 \cR -K_{,AA^*}(\p A\cdot\p A^*)+8(\p A)^2(\p A^*)^2(\cT+\cT^*)\notag\\
&+8(\p A)^2\Box A^*\phone+8(\p A^*)^2\Box A\phone^*\notag\\
&+\frac{16}3(\p A)^2(\p A^*)^2\big(K_{,A^*}\phone+K_{,A}\phone^*+4(A^{,m}\phone-A^{*,m}\phone^*)(A_{,m}\phone-A^*{}{,m}\phone^*)\big), \label{LagrangianBounce}
\end{align}
after elimination of all auxiliary fields. Here, as described in the Appendix, the notation $\cT,\phone$ refers to the lowest components of the Weyl-rescaled superfields $T,\Phone.$ Assuming now that $\cT=\cT^*$ and $\phone=\phone^*,$ and writing out the complex scalar $A$ in terms of two real scalars $\phi,\xi$ as
\be
A = \frac{1}{\sqrt{2}} (\phi + \I \xi),
\ee
the Lagrangian becomes
\begin{align}
\frac1{e}\mac{L}_{bounce}=&-\frac12 \cR -\frac12 K_{,AA^*}[(\p \phi)^2 + (\p \xi)^2] \notag\\ &+ [(\p \phi)^4 + (\p \xi)^4 - 2 (\p \phi)^2(\p \xi)^2 + (\p \phi \cdot \p \xi)^2]*[4 \cT + \frac{4}{3}\phone (K_{,A}+K_{,A^*})-\frac{32}{3}\phone^2 (\p \xi)^2]\notag\\
&+4 \sqrt{2}\phone[(\p \phi)^2\Box \phi - (\p \xi)^2 \Box \phi + 2 \p \phi \cdot \p \xi \Box \xi]\ . \label{LagrangianBounceReal}
\end{align}
Comparing to the Lagrangian \eqref{Lagrangian}, we can see that we should make the identifications
\bea
K_{,AA^*} &=& k(\phi), \\
\cT(\phi,\xi,\p \phi,\p \xi,\ldots) &=& \frac{1}{16} \tau(\phi), \label{eq:cT} \\ 
\phone(\phi,\xi) &=& -\frac{1}{8\sqrt{2}} g(\phi) \ .
\eea
Note that we have not included the term proportional to $\phone (K_{,A}+K_{,A^*})$ in equation \eqref{eq:cT} above. We easily could have done so, but it turns out that this term is negligibly small in the cases of interest to us, and therefore we are adopting a simpler definition of $\cT$ here. The above identifications can be realized by adopting the superfield definitions
\bea
K &=& - \frac12 (\Phi - \Phi^\dagger)^2 -\frac1{\kappa}(\Phi+\Phid)\arctan[\kappa(\Phi+\Phid)], \\
T &=& \frac{\bar{\tau}}{16}\frac{1}{(1+\kappa(\Phi + \Phid)^2)},\\
G &=& -\frac{\bar{g}}{8\sqrt{2}}\frac{1}{(1+\kappa(\Phi + \Phid)^2)} \ .
\eea
As we will see in section \ref{sectionstability}, further terms must be added to $K$ and $T$ in order for the scalar field $\xi$ to give rise to perturbations with a nearly scale-invariant spectrum and for it not to develop ghost and gradient instabilities during the bounce phase. These terms however only affect the perturbations of the $\xi$ field and are therefore irrelevant to the background dynamics. For completeness, we will write out the final form of the component Lagrangian during the bounce phase in terms of the two real scalars $\phi,\xi:$
\begin{align}
\frac1{e}\mac{L}_{bounce}=&-\frac12 \cR -\frac12 k(\phi)\left[(\p \phi)^2 + (\p \xi)^2\right] \notag\\ &+ \left[(\p \phi)^4 + (\p \xi)^4 - 2 (\p \phi)^2(\p \xi)^2 + (\p \phi \cdot \p \xi)^2\right] \times\notag\\
&\qquad \times\left[\frac{1}{4} \tau(\phi)+ \frac{1}{3}g(\phi)\left(\frac{\arctan(\sqrt{2}\kappa\phi)}{\sqrt{2}\kappa}+\frac{\phi}{1+2\kappa^2\phi^2}\right)-\frac{1}{12}g(\phi)^2 (\p \xi)^2\right]\notag\\
&- \frac12 g(\phi)\left[(\p \phi)^2\Box \phi - (\p \xi)^2 \Box \phi + 2 \p \phi \cdot \p \xi \Box \xi\right]\ , \label{LagrangianBounceRealFinal}
\end{align}
with the functions
\bea
k(\phi) &=& 1-\frac{2}{1 + 2 \kappa \phi^2},\\
\tau(\phi) &=& \frac{\bar{\tau}}{(1 + 2 \kappa \phi^2)^2}, \\
g(\phi) &=& \frac{\bar{g}}{(1 + 2 \kappa \phi^2)^2} \ .
\eea 
Regarding the terms in the third line of Eq. \eqref{LagrangianBounceRealFinal}, we note that the second term (proportional to $g(\phi)$) is everywhere at most a few percent of the magnitude of the first term (proportional to $\tau(\phi)$) since we are assuming $\bar{g} \ll \bar{\tau}$, while the last term (proportional to $g(\phi)^2$) is irrelevant to the background dynamics. This supergravitational extension of \eqref{Lagrangian} then reproduces the cosmic bounce described in section \ref{sectionmodel} above.

\subsection{The ekpyrotic phase}

In our model the bounce is preceded by an ekpyrotic contracting phase. During the ekpyrotic phase, the higher-derivative terms are effectively zero, $\cT,\phone \approx 0,$ and, as shown in the Appendix, the Lagrangian then reduces to
\begin{align}
\frac1{e}\mac{L}_{ekpyrotic}=&-\frac12 \cR -K_{,AA^*}(\p A\cdot\p A^*) - \e^K(K^{,AA^*}|D_AW|^2-3|W|^2) \, , \label{eq:Lagrangianekpyrotic}
\end{align}
where $D_A W = W_{,A} + K_{,A} W$ stands for the K\"{a}hler derivative of the superpotential. For an ekpyrotic phase, we need a potential $V$ that is steep enough and negative over a certain range of $\phi$. If the scalar potential is of the form
\be
V(\phi)=-V_0\e^{-c\phi}
\ee
then we need to have $c>\sqrt{6}$ as in this case the equation of state $w=p/\rho>1,$ which is the condition that is required for anisotropic stresses to be suppressed. To this end, we are considering a superpotential of the form
\be
W=\sqrt{V_0}w(A)\e^{-bA-\frac{dA}{A^2+1}} \ , \quad b,d\in\R \ , \quad b,d>0 \ , \label{eq:superpotential}
\ee
where the factor $w(A)$ has the property of being approximately equal to 1 for $\phi > \phi_{ek-end}$ while rapidly approaching zero for $\phi < \phi_{ek-end}$.\footnote{An alternative choice for the superpotential, with very similar properties, is $W(A)=\sqrt{V_0}w(A)A^f e^{-bA}$ with $0<f<\frac{1}{2}.$} Thus $w(A)$ ensures that the ekpyrotic phase comes to an end around $\phi_{ek-end}.$ As an example, we will take 
\be
w(A)=\frac12 \left[1+\tanh (\lambda(\sqrt{2}A-\phi_{ek-end})) \right] \,
\ee
with $\lambda$ being a positive real constant. For $\phi > \phi_{ek-end}$ we can simply approximate $w(A)\approx 1,$ and we will do so now in order to analyse the properties of the potential during the ekpyrotic phase.  We have written the second term in the exponent in \eqref{eq:superpotential} as $\frac{dA}{A^2+1}$ rather than the simpler choice $\frac{d}{A}$ in order to avoid a blow-up at $A=0.$ (Our superpotential thus has a pole at each of $A = \pm \I$. These are harmless since they have positive, and hence repulsive, potential energy and are located far from the vacuum region. If one prefers to avoid them, one can expand the factor $\frac{dA}{A^2+1}$ as a series and truncate it at the desired order, since we only require this approximate form over a certain field range.) For $d=0,$ the superpotential above would yield a potential 
\be
V\blc_{d=0}=V_0\e^{\xi^2-\sqrt2b\phi}\big(2\xi^2+b^2-3\big),
\ee
or, in the region of interest ($\xi=0$),
\be
V\blc_{\xi=0,d=0}=V_0\e^{-\sqrt2b\phi}\big(b^2-3\big).
\ee
For this potential we have $|V_{,\phi}/V|=\sqrt{2}b$ and $V<0$ for $b^2<3.$ Thus, one can easily see that this simple form for the superpotential allows either steep positive potentials, or shallow negative ones, exactly the opposite of what one is typically interested in in early universe cosmology. Indeed, an ekpyrotic phase requires $|V_{,\phi}/V|>\sqrt{6}$ and thus $b^2>3,$ which is in direct conflict with the requirement for negativity of the potential. It is here that the additional term proportional to $d$ is crucial. In fact, we will make the choice $b=\sqrt{3},$ which in the absence of $d$ would yield a vanishing potential. But when $d>0$ is turned on the potential becomes negative and sufficiently steep over a large (semi-infinite) field range, as can be seen from an explicit calculation of the potential:
\be
V\blc_{\xi=0}=\frac{4dV_0(\phi^2-2)}{(\phi^2+2)^4}\left( d(\phi^2 -2) -\sqrt{3}(\phi^2+2)^2\right) e^{-\sqrt{2}(\sqrt{3}+\frac{2d}{2+\phi^2})\phi} \ .
\ee
A straightforward calculation shows that the potential is negative for $\phi>\sqrt{2}$ when $d<16\sqrt{3},$ while for large $d$ it is negative when $\phi \gtrsim \sqrt{d}/3^{1/4}.$ Moreover, the potential is sufficiently steep, i.e. $|V_{,\phi}/V|>\sqrt{6}$ for $\phi > \sqrt{2} d$ when $d \gtrsim 2,$ while for small $d$ the potential is sufficiently steep for $\phi \gtrsim 2.5$. Thus, for all $d>0$ this potential is suitable for an ekpyrotic phase of arbitrarily long duration, as long as $\phi \gtrsim \{ 2.5, \sqrt{2}d \},$ whichever happens to be the stronger condition. Since we are mostly interested in the field range where $\phi$ is large, $\phi \gtrsim 10,$ and $\xi$ near zero, the potential can be well approximated by 
\be
V\blc_{\phi \gg 1, \xi=0}=-\frac{4\sqrt{3}dV_0}{\phi^2} e^{-\sqrt{6}\phi-2\sqrt{2}\frac{d}{\phi}}.
\ee
\begin{figure}
\includegraphics[width=0.75 \textwidth]{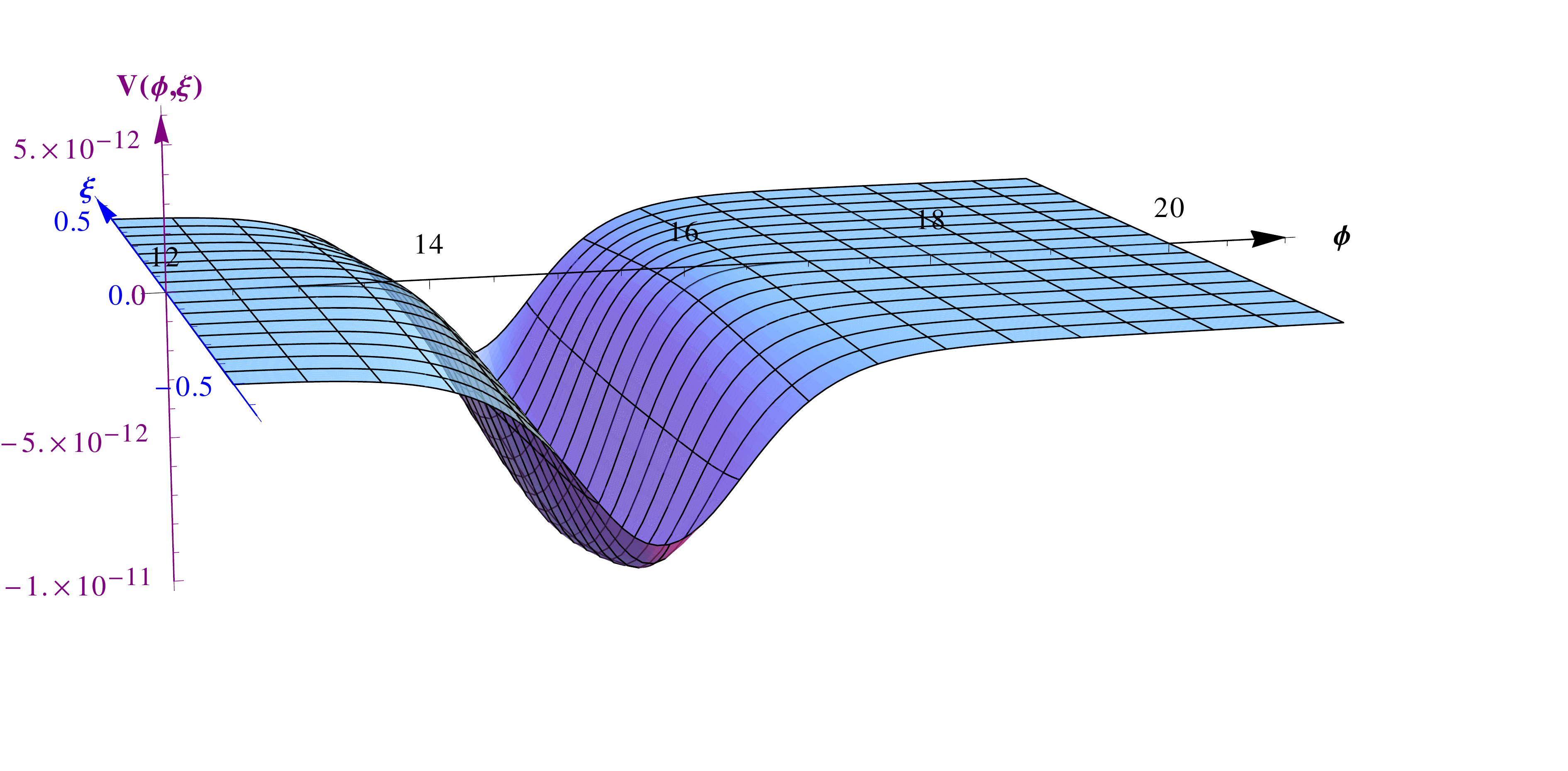}
\caption{The ekpyrotic potential corresponding to the superpotential \eqref{eq:superpotential} in the vicinity of the $\xi=0$ line, for $b=\sqrt{3}$ and $d=1$. The ekpyrotic phase occurs at large, decreasing, $\phi$ and comes to an end near $\phi_{ek-end}\sim 15.$ Then the potential turns off and the kinetic/bounce phase ensues. In the transverse $\xi$ direction, the potential is essentially flat (closer inspection would reveal a small positive curvature).}\label{fig:FullPot}
\end{figure}
Reinstating the factor $w(A),$ we can now plot the full potential -- see Fig. \ref{fig:FullPot} for an example with $d=1.$ In the figure, the ekpyrotic phase comes to an end near $\phi_{ek-end}\sim 15.$ The potential then comes back up to zero and the kinetic/bounce phase follows. We should note that for certain larger values of $d$ the potential slightly overshoots as it comes back up to zero for $\phi \lesssim \phi_{ek-end},$ leading to a small positive bump. This is caused by the derivatives in the formula for the potential \eqref{eq:Lagrangianekpyrotic} acting on the $w(A)$ factor in the superpotential (also see the discussion in \cite{Lehners:2002tw}). However, such a small bump has no noticeable effect on the dynamics, as the evolution is completely dominated by the kinetic energy of $\phi$ at that stage and so the field simply flies over the bump\footnote{It is useful to bear in mind that the universe is contracting during this phase, so that kinetic energies get significantly blue-shifted. Falling 
down the ekpyrotic potential and climbing out of it again are two highly asymmetric evolutions.}. Thus, we have now demonstrated that an ekpyrotic phase can be embedded into our ${\cal N}=1$ supergravity model, and, moreover, that the ekpyrotic phase turns off smoothly in preparation for the bounce phase.

%%%%%%%%%%%%%%%%%%%%%%%%%%%%%%%%%%%%%%%%%%%%%%%%%%%%%%%%%%%%%%%%%%%%%%%%%%%%%%%%%%%%%%
\section{Perturbations and Stability} \label{sectionstability}

In order to verify the trustworthiness of the background ekpyrotic plus bounce evolution that we have described up to now, we must check that our model does not contain dangerous instabilities. The most important such criterion is the absence of perturbative ghost instabilities -- that is, we must ensure that quadratic perturbations of the kinetic terms around our background solution contain correct-sign time-derivative terms (implying that time-dependent perturbations cost kinetic energy as opposed to releasing kinetic energy). Otherwise, our background solution would be catastrophically unstable. This is easily achieved in our theory. A second criterion is that spatial gradient perturbations should also have the correct sign in order to avoid gradient instabilities, at least over most of the evolution. Gradient instabilities result in the growth of perturbations, and if they go unchecked they signal the breakdown of our description. As we will see, during the bounce phase there is a brief period over which 
gradient instabilities of the $\phi$ field are present, as was already described in \cite{Cai:2012va}. However, because this period is very brief, it simply results in a small growth of the perturbations, and is not dangerous as such. 

What is really new to our analysis here is that we must also ensure the stability of the second scalar field $\xi$ that is required by supersymmetry. It turns out that the stability of $\xi$ is non-trivial, in the sense that it requires us to include an additional stabilizing term to the original action. We present a simple example of a stabilizing term, which ensures the stability of $\xi$ throughout the bounce phase.

Finally, one may wonder if the model we are presenting can also generate primordial density perturbations in agreement with observations of the cosmic background radiation. This can indeed occur via the so-called entropic mechanism, where nearly scale-invariant entropy perturbations (which here correspond to perturbations in the second scalar $\xi$) are generated during the ekpyrotic phase, and get converted into curvature perturbations during the short kinetic phase preceding the bounce. We will present the detailed conditions for this to happen below. 

\subsection{Absence of ghosts}

\subsubsection{Stability of $\phi$}

\begin{figure}
\includegraphics[width=0.75 \textwidth]{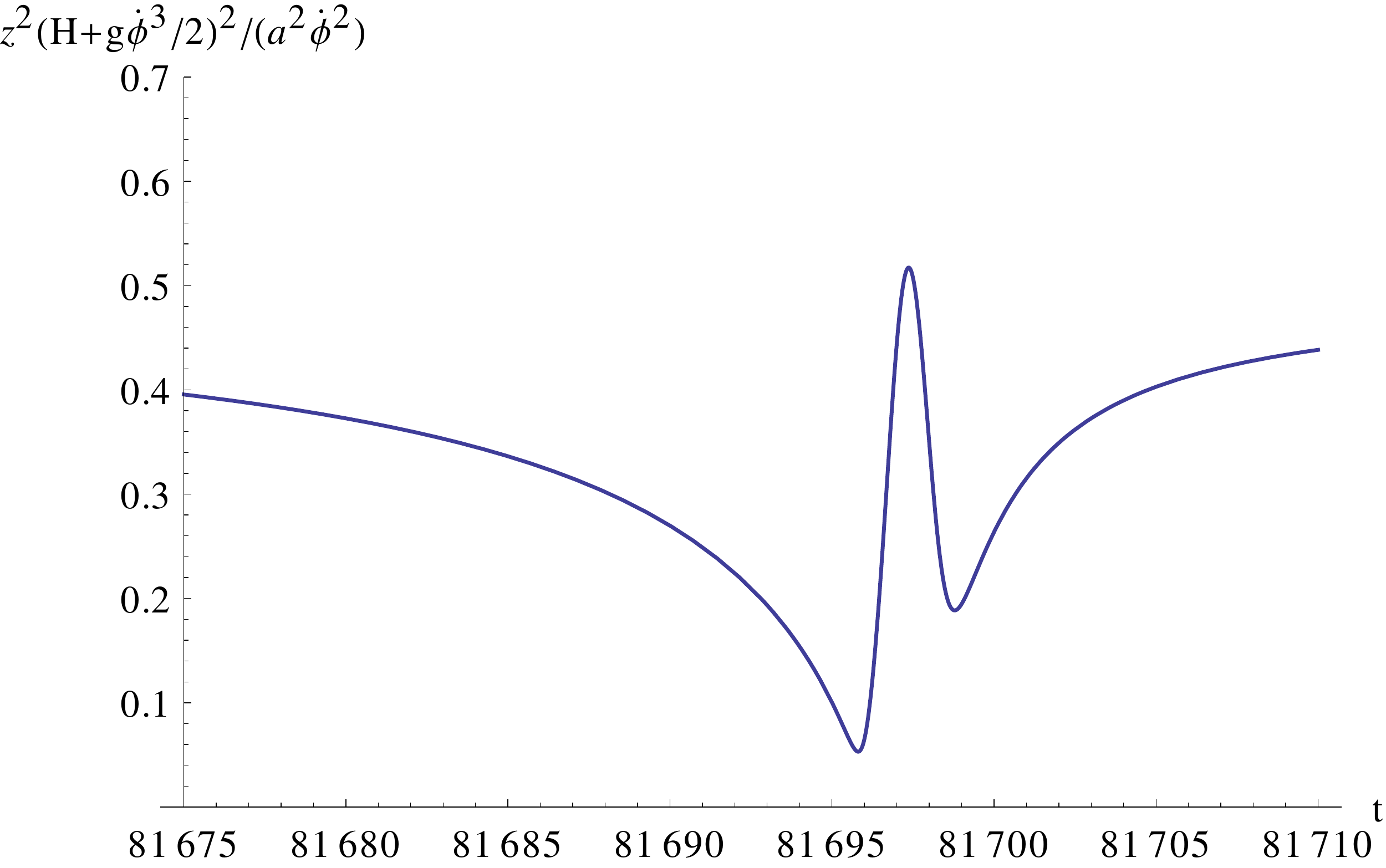}
\caption{This plot shows the evolution of the quantity $z^2 (H+\frac12 g \dot\phi^3)^2/(a^2 \dot\phi^2)$ over the time of the bounce. The positivity of this quantity ensures the absence of ghost instabilities of scalar curvature perturbations.}\label{fig:PhiTime}
\end{figure}

Perturbations in the scalar field $\phi$ are not gauge-invariant and hence, in order to discuss the stability of the scalar field driving both the ekpyrotic and the bounce phases, we must look at the stability of the curvature perturbation $\zeta$ (which corresponds to a local gauge-invariant space-time-dependent perturbation of the scale factor). As derived in \cite{Deffayet:2010qz,Gao:2011qe}, the quadratic action for $\zeta$ is given by
\be
S_{(2)} \supset \int dt d^3x a z^2 \left[ \dot\zeta^{2} - \frac{c_s^2}{a^2} (\p_i \zeta)^2 \right] \label{eq:zetaquad}
\ee
with 
\bea
z^2 &=& \frac{a^2 \dot\phi^2}{(H+\frac12 g \dot\phi^3)^2} \left( \frac12 k + \frac32 \tau \dot\phi^2 -3gH\dot\phi +\dot{g}\dot\phi +\frac34 g^2 \dot\phi^4 \right), \label{eq:z2} \\\label{eqn:cs2}
c_s^2 &=& \frac{k + \tau \dot\phi^2 -4gH\dot\phi -2g\ddot\phi-\frac12 g^2 \dot\phi^4}{k + 3 \tau \dot\phi^2 -6gH\dot\phi +2\dot{g}\dot\phi +\frac32 g^2 \dot\phi^4},
\eea
where $c_s^2$ has the physical interpretation of being the square of the speed of propagation of the fluctuations. The absence of ghosts corresponds to the requirement that $z^2$ be positive throughout. The fraction in the definition of $z^2$ above is evidently positive, and so the crucial requirement is that the expression in the large parentheses be positive. As shown in Fig. \ref{fig:PhiTime}, our example easily satisfies this criterion (we are only plotting the period of the bounce, which is the only time when the question of stability is non-trivial), and thus the curvature perturbations remain ghost-free throughout.

\begin{figure}
\includegraphics[width=0.75 \textwidth]{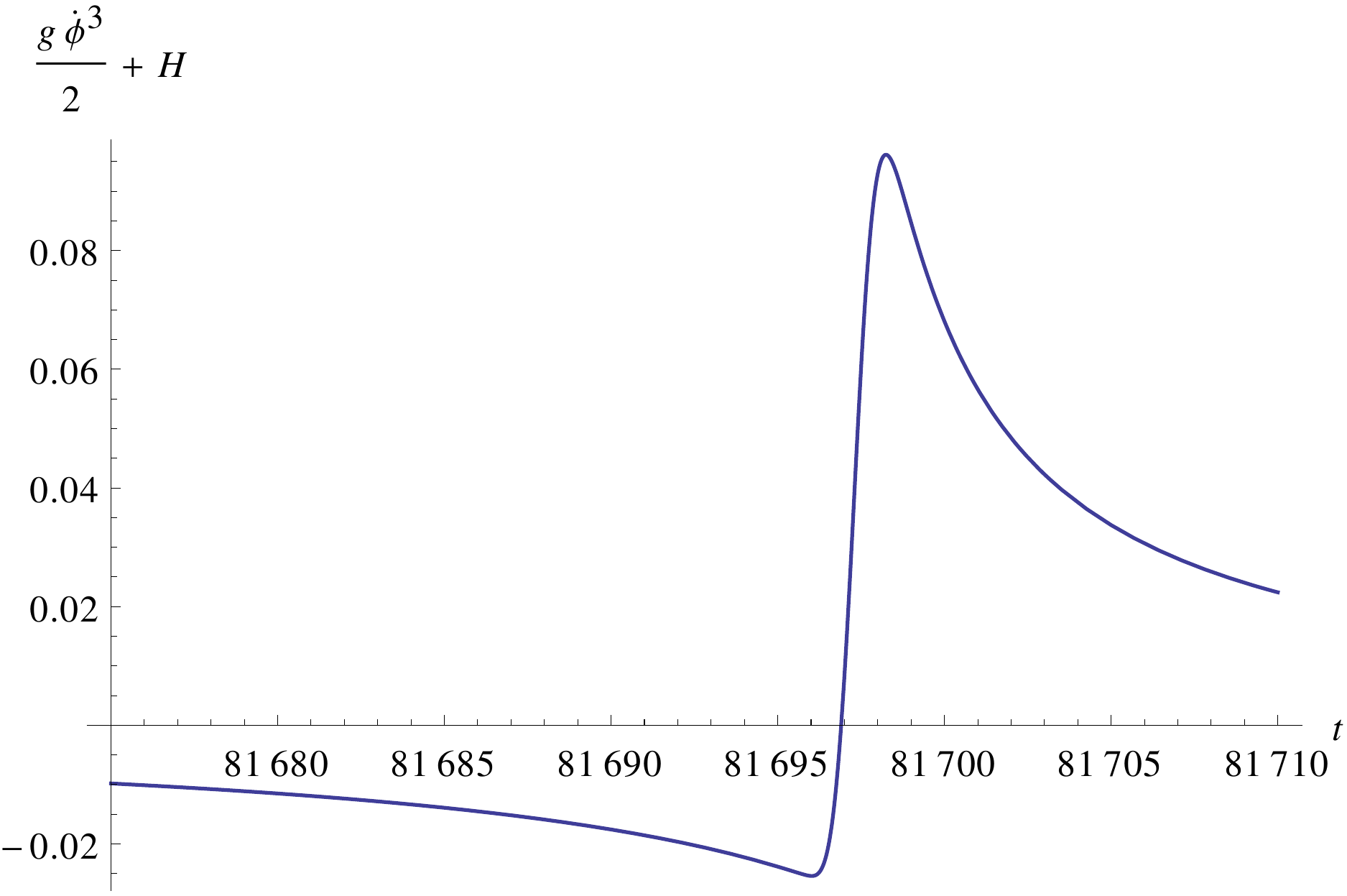}
\caption{This plot shows the evolution of the quantity $H+\frac12 g \dot\phi^3$ over the time of the bounce.}\label{fig:HplusGalileon}
\end{figure}

\begin{figure}
\includegraphics[width=0.75 \textwidth]{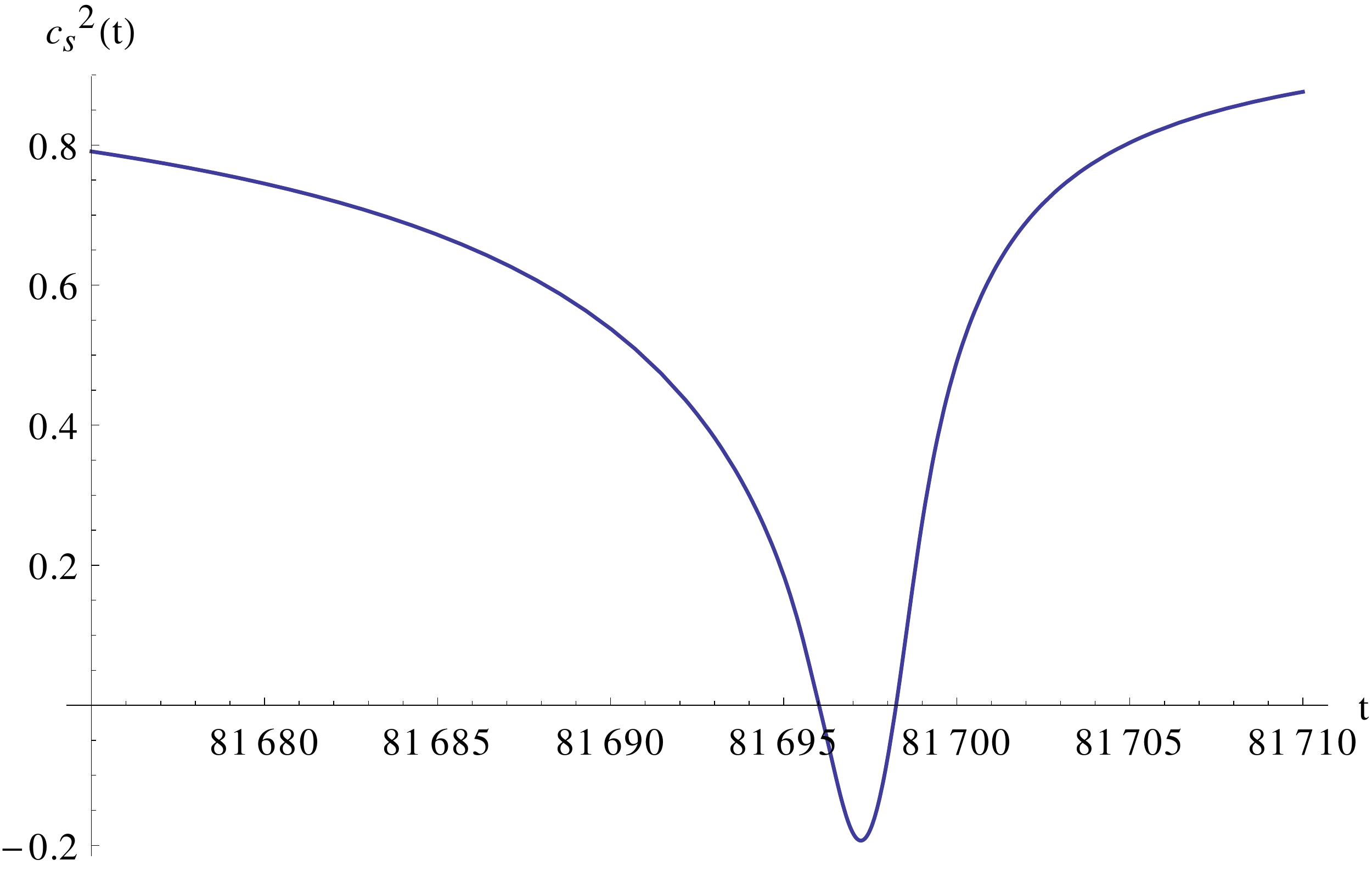}
\caption{This plot shows the evolution of the square of the speed of sound $c_s^2$ during the bounce phase. A brief period of instability arises when $c_s^2$ becomes negative -- this causes an extra growth of the curvature perturbation $\zeta.$}\label{fig:PhiSpace}
\end{figure}

Likewise, positivity of the square of the speed of sound $c_s^2$ is associated with the stability of gradient perturbations. A plot of $c_s^2$ is shown in Fig. \ref{fig:PhiSpace}. As can be seen, during the bounce phase, there is a brief period during which $c_s^2$ becomes negative, signalling a brief instability. Comparing with Fig. \ref{fig:NEC}, we can see that the instability is coincident with the period of NEC violation.
%%% CHANGES BY MK 14-1-22:
From \eqref{eqn:cs2}, we can furthermore see that away from the bounce the speed of sound tends to unity from below, such that superluminality as an obstruction to UV completion can be excluded (cf.~\cite{Adams:2006sv}). The numerical solution of the equation of motion
\be\label{eom-zeta}
\ddot{\zeta} + \left( H + 2\frac{\dot{z}}{z}\right) \dot\zeta + \frac{c_s^2 k^2}{a^2}\zeta  =0 \ ,
\ee
for the curvature perturbations $\zeta$ is problematic in this gauge because the denominator $(H+\frac12 g \dot\phi^3)^2$ appearing in \eqref{eq:z2} will necessarily become zero in the vicinity of the bounce (cf.~Fig.~\ref{fig:HplusGalileon}), and thus \eqref{eom-zeta} momentarily becomes singular as $H+\frac12 g \dot\phi^3$ passes through zero.
This singularity shows that the gauge used here (namely constant scalar field gauge \cite{Gao:2011qe}) is not the appropriate choice in the vicinity of the bounce. This is despite the fact that $\dot\phi\neq0$ for the whole evolution, cf.~Figs.~\ref{fig:Phi} and \ref{fig:PhiDot}. While it is important to make sure that perturbations of small wavelengths do not destroy the homogeneous background evolution during the bounce phase, the observed singularity in \eqref{eom-zeta} is not in itself problematical -- such difficulties generically arise when perturbing about a solution where a background quantity goes through an extremum. It merely shows that one should work in a gauge where the perturbation equations remain well-defined throughout. An example for such a suitable gauge choice would be the harmonic gauge. By use of the harmonic gauge, the authors of \cite{Xue:2013bva} recently carried out a full numerical study in a simpler (and non-supersymmetric) bounce model, and they found that the curvature 
perturbation indeed grows by a small amount across the bounce phase.\footnote{Note that $\dot\phi$ goes through zero momentarily in the model of \cite{Xue:2013bva}, as opposed to our case.} We expect similar results to hold in our case and plan on presenting an analogous analysis adapted to our model in forthcoming work.
%%% ORIGINAL from v2:  As previously discussed in \cite{Cai:2012va}, this instability causes the curvature perturbations on large scales not to remain conserved, and in fact causes the curvature perturbation to acquire an extra growth. This additional growth of the amplitude of $\zeta$ is however limited by the brevity of the unstable period. In \cite{Cai:2012va} the equation of motion for $\zeta,$ written in Fourier space as
%%%\be
%%%\ddot{\zeta} + \left( H + 2\frac{\dot{z}}{z}\right) \dot\zeta + \frac{c_s^2 k^2}{a^2}\zeta  =0 \ ,
%%%\ee
%%% was solved numerically in order to estimate this growth. We prefer not to do so here, as the denominator $(H+\frac12 g \dot\phi^3)^2$ appearing in \eqref{eq:z2} necessarily becomes zero in the vicinity of the bounce, and thus the equation momentarily becomes singular. This shows that the gauge used here (namely constant scalar field gauge \cite{Gao:2011qe}) becomes ill-defined in the vicinity of the bounce. This fact is not in itself problematical -- such difficulties generically arise when perturbing about a solution where a background quantity goes through an extremum. It just means that one has to work in a gauge where the perturbation equations remain well-defined throughout. An example of such a well-defined gauge is harmonic gauge, which was also recently used in \cite{Xue:2013bva}, where the authors carried out a full numerical study in a simpler (and non-supersymmetric) bounce model, and where they found that the curvature perturbation indeed grows by a small amount across the bounce phase. We 
expect similar results to hold in our case and plan on presenting an analogous analysis adapted to our model in forthcoming work.

\begin{figure}
\includegraphics[width=0.75 \textwidth]{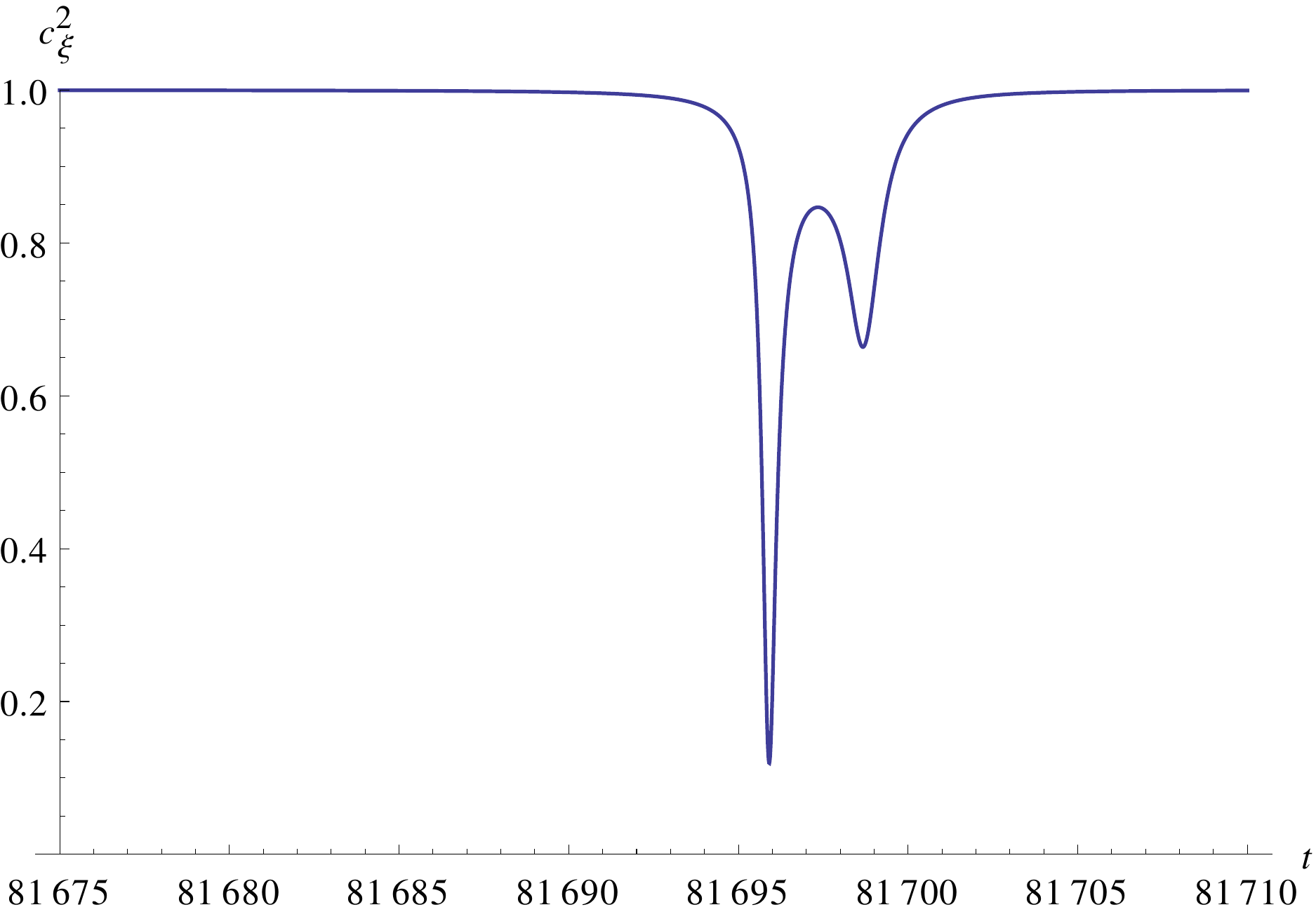}
\caption{This plot shows the evolution of the square of the speed of sound $c_\xi^2$ of the second scalar $\xi$ after inclusion of the stabilizing term \eqref{eq:stabilizing}.}\label{fig:cxi2}
\end{figure}

\subsubsection{Stability of $\xi$} \label{sectionstabilityxi}

\begin{figure}
\includegraphics[width=0.75 \textwidth]{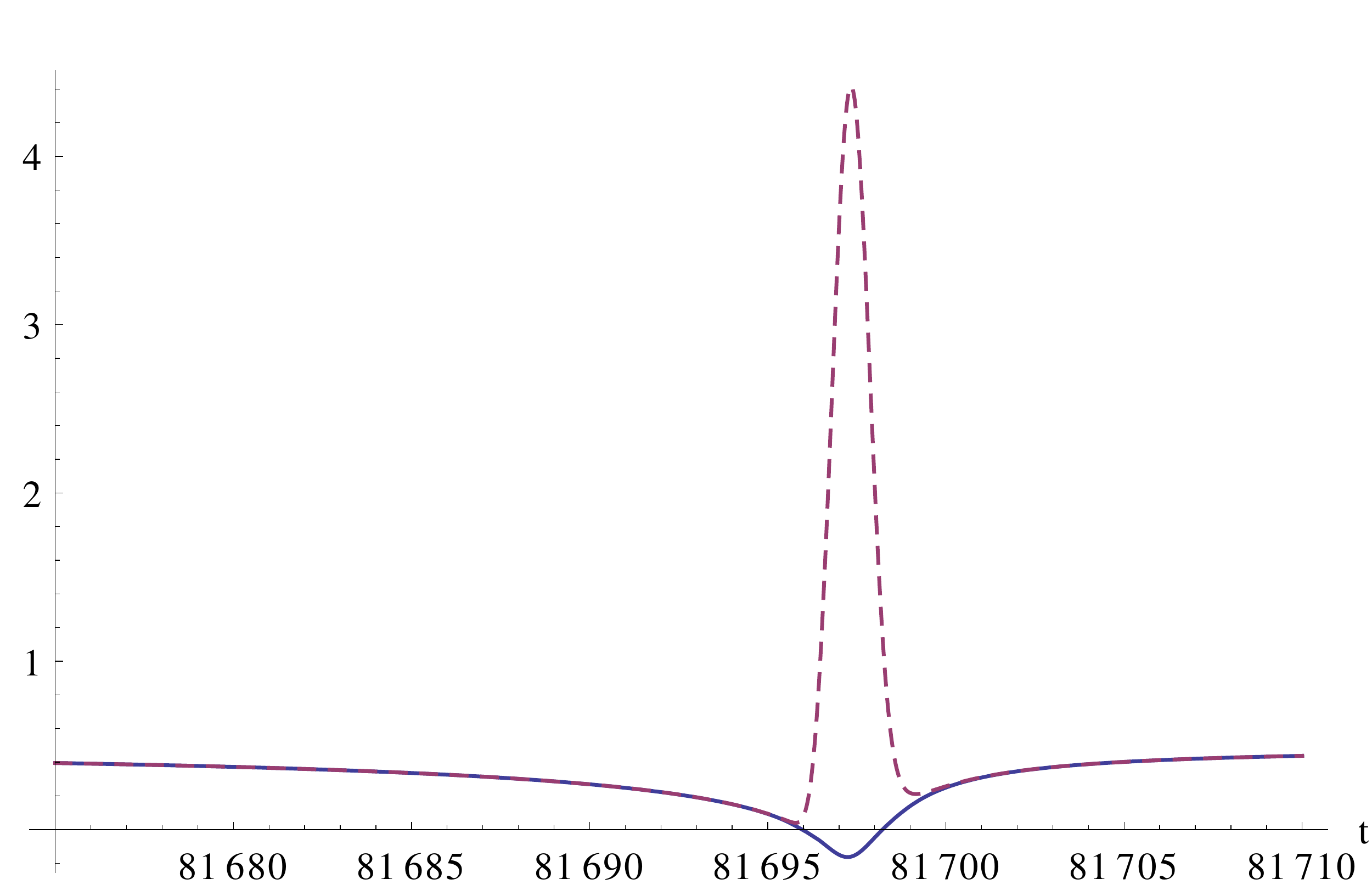}
\caption{A plot of the coefficient of time-dependent fluctuations $(\dot{\delta \xi})^2$ in the perturbed Lagrangian. Positivity ensures the absence of ghost instabilities. The solid curve indicates that our original Lagrangian contains ghost instabilities, but with the inclusion of a stabilizing term \eqref{eq:stabilizing} these ghosts are avoided (dashed curve, a zoom-in would confirm strict positivity throughout).} \label{fig:XiTime}
\end{figure}

\begin{figure}
\includegraphics[width=0.75 \textwidth]{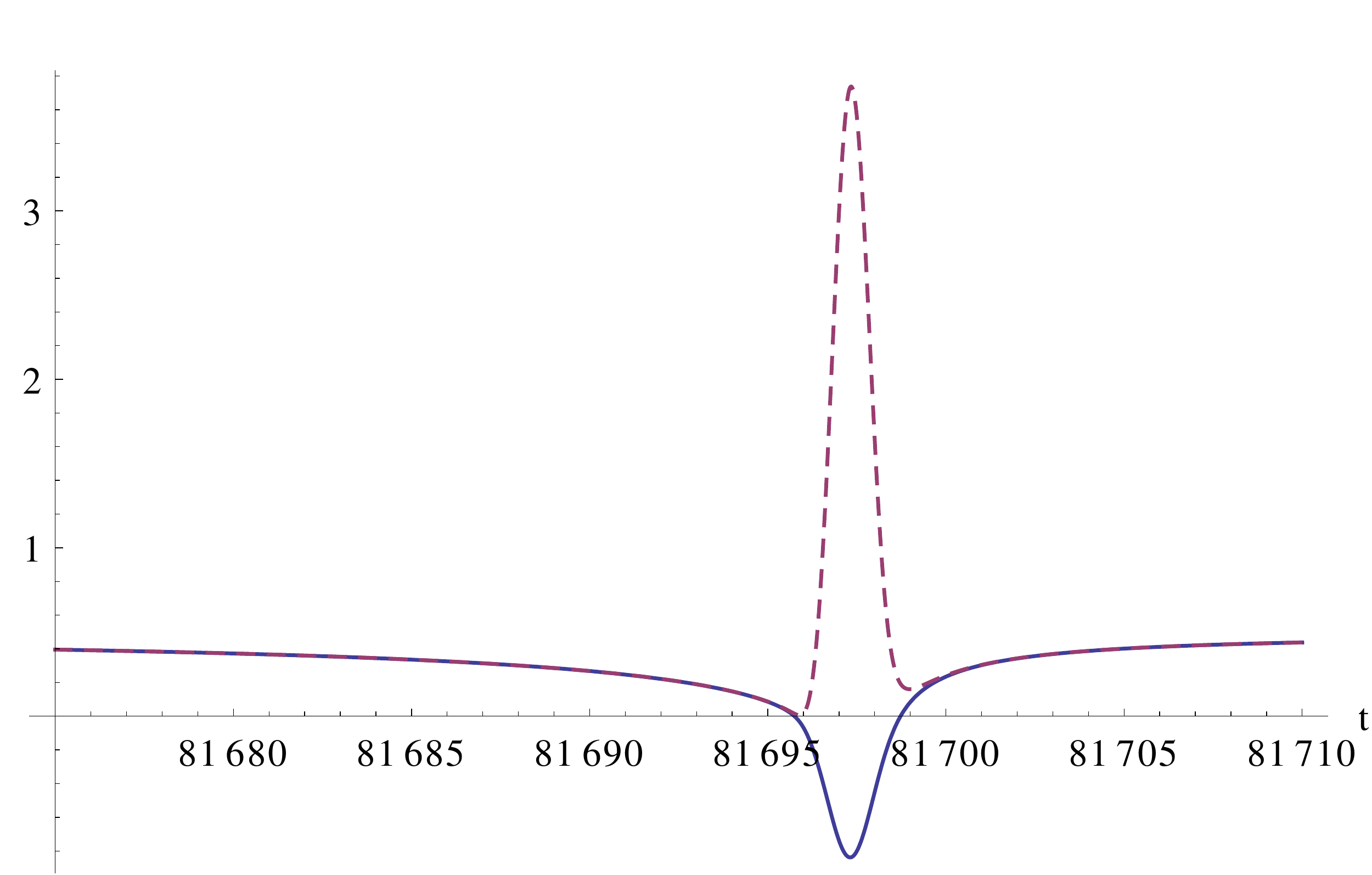}
\caption{A plot of the coefficient of space-dependent fluctuations $({\delta \xi_{,i}})^2$ in the perturbed Lagrangian. Positivity ensures the absence of gradient instabilities. The solid curve indicates that our original Lagrangian contains gradient instabilities, but with the inclusion of a stabilizing term \eqref{eq:stabilizing} these instabilities are avoided (dashed curve, a zoom-in would confirm strict positivity throughout). }\label{fig:XiSpace}
\end{figure}

Supersymmetry requires the presence of a second real scalar field $\xi.$ Although this field does not contribute to the background dynamics in our model, its fluctuations are nevertheless of crucial importance. Indeed, we must verify under what conditions this second scalar can destabilize the model that we have presented so far. In order to do so, we must calculate the action for $\xi$ up to quadratic order in fluctuations. In the present section, we will discuss the associated kinetic and gradient terms, while in the next section we will discuss the stability properties of the scalar potential in the $\xi$ direction. Since $\xi$ is a scalar field pointing transverse to the background trajectory in scalar field space, its fluctuations are automatically gauge-invariant and they correspond to entropy/isocurvature perturbations \cite{Gordon:2000hv}. We find that the quadratic action for the time- and space-derivatives of these fluctuations (i.e. ignoring for now the perturbations in the mass of the 
fluctuations) is given by
\bea
S_{(2)} &\supset& \int dt d^3 x \left[ a^3 (\dot{\delta \xi})^2 \left( \frac12 k +\frac12 \tau \dot\phi^2 + g \ddot\phi +\frac12 \dot{g}\dot\phi +\frac{1}{12}g^2 \dot\phi^4 \right) \right.\nn\\ 
&& \qquad \quad \left. -a (\p_i \delta\xi)^2 \left( \frac12 k -\frac12 \tau \dot\phi^2 + g H\dot\phi -\frac12 \dot{g}\dot\phi +\frac{1}{12}g^2 \dot\phi^4\right) \right] \ .
\eea
Stability is synonymous with the terms in parentheses being positive. As things stand, this requirement is not satisfied -- see the solid curves in Figs. \ref{fig:XiTime} and \ref{fig:XiSpace}. However, we can easily extend our model to include a suitable stabilizing term, without affecting the background dynamics.  
For instance, consider adding the following term to the higher-derivative coefficient function $T:$
\be
\Delta T = \frac{1}{32} c_\xi \tau(A,A^*) \p^m(A-A^*) \p_m(A^*-A),
\ee
where $c_\xi$ is a real and positive constant. This adds a contribution
\be
\frac1e \Delta {\cal L}= - c_\xi \tau(\phi) \dot\phi^4 (\p\xi)^2 \label{eq:stabilizing}
\ee
to the full component Lagrangian. During the ekpyrotic phase, this term is negligibly small, both because $t(\phi)\approx 0$ at that stage and due to the smallness of $\dot\phi^4$. However, during the bounce phase we obtain significant additional contributions to the time and space derivative fluctuation terms. In our explicit example, we have found that a value $c_\xi=1$ is sufficient to ensure the absence of ghost instabilities during the bounce phase, which is a sufficient criterion for stability if one is prepared to tolerate a brief growth of the $\xi$ perturbations due to gradient instabilities. Otherwise, a larger value of $c_\xi \approx 10$ eliminates both ghost and gradient instabilities -- see the dashed curves in Figs.  \ref{fig:XiTime} and \ref{fig:XiSpace}. We note that we have provided merely one example of such a stabilizing term - many other terms would achieve the same effect.

We should add a comment about the stability properties of the Galileon term proportional to $g(\phi).$ As demonstrated in our earlier paper \cite{Koehn:2013hk}, in global supersymmetry this term leads to ghost instabilities around non-trivial $\xi$ backgrounds and/or backgrounds with large $\ddot\phi$ contributions, and the best one can achieve is to have a perturbatively stable background in an effective field theory context. What we are showing here is that when coupled to supergravity, one can indeed ensure the absence of {\it perturbative} ghost instabilities. More specifically, with the inclusion of a stabilizing term such as the one presented in \eqref{eq:stabilizing}, ghost perturbations are entirely avoided in our model.

\subsection{Entropy perturbations}

Up to now, we have described a model that allows for the universe to bounce in a non-singular and controlled manner, without catastrophic instabilities, and with a prior ekpyrotic phase ensuring that the universe enters the bounce phase with near-perfect spatial homogeneity and isotropy. Thus, we have provided a viable model for a bounce in supergravity, irrespective of the question of the origin of the primordial density/temperature fluctuations seen in the cosmic background radiation. These could for example arise during an inflationary period after the bounce (in which case we have provided a possible pre-history to inflation, which might be able to explain some of the large-scale anomalies seen in the CMB \cite{Piao:2003zm,Biswas:2013dry,Liu:2013kea}) or during a matter contraction phase before the ekpyrotic phase as envisaged in \cite{Cai:2013kja}. Here, we are interested in the question whether we can augment our model so as to allow for the generation of nearly scale-invariant density perturbations 
directly during the ekpyrotic phase. (Note that the present section is independent of the rest of the paper.) 

The currently best-known way in which this can be achieved is via the entropic mechanism \cite{Lehners:2007ac,Buchbinder:2007ad,Buchbinder:2007tw,Buchbinder:2007at,Creminelli:2007aq}, where scale-invariant entropy perturbations are amplified first, and are subsequently converted into curvature perturbations during the kinetic phase preceding the bounce \cite{Battarra:2013cha}. A conversion can be achieved if the trajectory in scalar field space bends. Such a bend could easily be incorporated into our model by adding an effective potential during the kinetic phase -- see \cite{Lehners:2006pu,Lehners:2007nb} for a derivation of such a potential. Alternatively, conversion can occur during reheating at the bounce \cite{Battefeld:2007st}. What is really crucial for the model to work is that the entropy perturbations have to acquire the correct spectrum. The spectrum is related to the transverse curvature of the potential, as we will now review.  During the ekpyrotic phase, the equation of motion for the entropy 
perturbations is given at linear order (and in Fourier space) by
\be
\ddot{\delta \xi} + 3 H \dot{\delta \xi} + \left( \frac{k^2}{a^2} + V_{,\xi\xi} \right) \delta \xi = 0.
\ee
In terms of the re-scaled variable $\delta \Xi = a \delta \xi$ and in terms of conformal time $d\tau = d t /a,$ this becomes
\be
\delta \Xi^{\prime\prime} + \left( k^2 - \frac{a^{\prime\prime}}{a} + a^2 V_{,\xi\xi} \right) \delta \Xi = 0.
\ee
Defining $V \equiv V(\phi) \left( 1 - m_\xi^2 \xi^2 + \cdots \right),$ and using the ekpyrotic background solution \eqref{eq:ek} we obtain
\be
\frac{a^{\prime\prime}}{a} - a^2 V_{,\xi\xi} = \tau^2 \left( -\frac{\epsilon -2}{(\epsilon -1)^2} - m_\xi^2 \frac{\epsilon - 3}{\epsilon^2} \right),
\ee
where $\epsilon$ is related to the equation of state $w,$ which was defined in \eqref{eq:w}, via the usual relation $\epsilon = \frac{3}{2}(1+w).$ A standard calculation then shows that the solutions (given in terms of Hankel functions) to this equation are fluctuation modes with a spectral index
\be
n_s = 4 - 2 \nu \ ,
\ee
where $\nu$ is given by ($\nu$ is the index of the Hankel function in question)
\be
\nu^2 = \frac{1}{4} -\frac{\epsilon -2}{(\epsilon -1)^2} - m_\xi^2 \, \frac{\epsilon - 3}{\epsilon^2}.
\ee
Thus, if we want to obtain a nearly scale-invariant spectrum, $n_s \approx 1,$ we must have $\nu^2 \approx 9/4$ and this in turn requires (assuming $\epsilon$ to be very close to, but a little bigger than, $3$)
\be
m_\xi^2 \approx -\frac{81}{4(\epsilon - 3)}.
\ee
This shows that the potential must be negatively curved in the transverse direction. That is, over the field range where the modes of observational interest are generated, the potential must be tachyonic. The possible implications of this fact have been discussed in detail in \cite{Lehners:2008qe,Lehners:2009eg}. 

We now have to compare this requirement with the transverse curvature of the potential that we have been using up to now. Starting from \eqref{eq:superpotential}, a straightforward calculation shows that at large $\phi$ and up to quadratic order in $\xi$ the potential is given by
\be
V\blc_{\phi \gg 1, {\cal O}(\xi^2)} \approx V_0 e^{-\sqrt{6}\phi-2\sqrt{2}\frac{d}{\phi}} \left( -\frac{4\sqrt{3}d}{\phi^2} +\frac{4d^2}{\phi^4} +\xi^2 \left( 2 -\frac{2\sqrt{3}d}{\phi^2}\right)\right),
\ee
where we have written out only the leading terms. This implies an effective mass for $\xi$ given by
\be
m_{\xi}^2 = -\frac{V_{,\xi\xi}}{V} \blc_{\phi \gg 1, \xi=0} \approx \frac{\phi^2}{\sqrt{3}d} \left( 1-\frac{2d}{\sqrt{3}\phi^2} \right),
\ee
which in fact corresponds to a stable mass term. Thus, as it stands, our potential leads to a stable ekpyrotic phase but not to the generation of a scale-invariant spectrum of entropy perturbations. However, we can consider adding an additional term to the K\"{a}hler potential, of the form
\be
\Delta K = \frac14 (A-A^*)^4 p\left( \frac{1}{\sqrt{2}}(A+A^*)\right)= \xi^4 p(\phi)
\ee
so that $K_{,AA^*}$ is augmented by a term $6\xi^2 p(\phi).$ This additional term does not affect any of our preceding analysis, but leads to a change in the second derivative of the potential given by
\be
\Delta V_{,\xi\xi}\blc_{\phi \gg 1, \xi=0} \approx  e^K{K^{,AA^*}}_{\xi\xi}|D_A W|^2 =  -12 p(\phi) V_0 e^{-\sqrt{6}\phi-2\sqrt{2}\frac{d}{\phi}} \left( 3-\frac{4\sqrt{3}d}{\phi^2} \right).
\ee
Thus the total transverse mass squared now becomes (to leading order at large $\phi$)
\be
m_\xi^2 = \frac{V_{,\xi\xi}}{V} \blc_{\xi=0} \approx \frac{1- 9 p(\phi)}{\sqrt{3} d}\phi^2
\ee
Putting all of these results together, we can see that a scale-invariant spectrum can be obtained for
\be
9 p(\phi)  \approx \frac{81\sqrt{3}d}{4\phi^2(\epsilon - 3)} + 1
\ee
Thus, from a model-building perspective, we can design our model such that a nearly scale-invariant spectrum of perturbations is obtained. 

We note that the relation above only needs to be satisfied over the range of $\phi$ where the modes of observational interest are being generated. Since in our model $\epsilon \approx 3,$ we have the relationship that a change $\Delta {\cal N}$ in the number of e-folds is related to a change of $\phi$ via $\Delta \phi \approx 1.2 \Delta {\cal N}.$ Hence the modes of observable interest (which are generated between about $50$ and $60$ e-folds before the end of the ekpyrotic phase) correspond to $\phi$ in the range $\phi_{ek-end}+60$ to $\phi_{ek-end}+75.$ At larger $\phi,$ it would in fact be desirable if $p(\phi)$ approached zero again, since this would make the potential stable at the beginning of ekpyrosis. 

Evidently, simply adding the required term as above is highly tuned. What is largely responsible for the required amount of tuning is the fact that the potential for $\phi$ is only just steep enough for an ekpyrotic phase. For steeper potentials (of the form $V(\phi)=-V_0 e^{-c\phi}$ with $c\gtrsim 10$) the simple relationship $V_{,\phi\phi}\approx V_{,\xi\xi}$ would guarantee a spectral index that is close to scale-invariant, with deviations from scale-invariance of order $1/c^2$ and thus at the percent level \cite{Lehners:2013cka}. Here, however, we have $c^2 \approx 6,$ and thus deviations from scale-invariance are typically rather large, and substantial fine-tuning is required in order to obtain a spectrum in agreement with observations. Thus, in the present context, the entropic mechanism for producing density perturbations appears rather unnatural. 

An important question would therefore be to see if a tachyonic mass of the form required here could arise in a more natural manner, perhaps from an axion field residing near a maximum of its potential. Another possibility is to consider variants of the entropic mechanism for producing density perturbations. Particularly promising is the recently proposed model with a non-minimal coupling between the two scalars $\phi$ and $\xi$ \cite{Qiu:2013eoa,Li:2013hga}. In this model, no unstable potential for $\xi$ is required. Incorporating this model into supergravity, and combining it with a bounce, is currently work in progress. 

%%%%%%%%%%%%%%%%%%%%%%%%%%%%%%%%%%%%%%%%%%%%%%%%%%%%%%%%%%%%%%%%%%%%%%%%%%%%%%%%%%%%%%%%
\section{Discussion} \label{sectiondiscussion}

Obtaining a viable, stable model for a non-singularly bouncing universe is non-trivial since in a flat universe the null energy condition must be violated during the bounce phase. No type of matter is currently known which can achieve this. However, near the Big Bang currently experimentally validated physical theories break down and we know that new physics must come into play. In string theory, for example, new types of matter are predicted to play a fundamental role, including in particular negative-tension branes \cite{Horava:1995qa,Lukas:1998yy,Lukas:1998tt,Donagi:1998xe,Braun:2005nv,Lehners:2007nb}. These contain negative energy density, and thus might be able to lead to effective violations of the null energy condition from the four-dimensional point of view. This raises the question of whether it is conceivable that non-singular bounces can occur in nature.

In the present paper, we have provided an argument for answering this question in the affirmative. We have considered a supergravitational version of scalar field theories with higher-derivative kinetic terms, of a form that may arise in the dynamical description of branes \cite{deRham:2010eu,Khoury:2012dn,Ovrut:2012wn}. More specifically, we have made use of ghost condensate and Galileon theories, and in this context we have shown that it is possible to construct a stable non-singular bounce model. This result is entirely non-trivial, since one might have expected that the stability and rigidity associated with supergravity theories would not have allowed NEC-violating, yet perturbatively stable, solutions. Given that supergravity theories are expected to be good approximations to string theory at the energy scales that are relevant here (i.e. energy scales a few orders of magnitude below the full quantum gravity scale), our results provide an indication that non-singular bounces are indeed allowed in 
string theory. 

Our proof-of-principle that non-singular bounces exist in supergravity raises interesting issues, especially in the context of cosmology. On the one hand, it lends further credence to ekpyrotic/cyclic models \cite{Khoury:2001wf,Steinhardt:2001st,Lehners:2008vx} as viable alternatives to inflationary models (with predictions in good agreement with data, plus a number of conceptual advantages \cite{Lehners:2013cka}). On the other hand, the existence of cosmic bounces dramatically changes the predictions in a landscape context -- see \cite{Piao:2004me,Piao:2009ku,Johnson:2011aa,Lehners:2012wz,Garriga:2012bc,Garriga:2013cix,Gupt:2013poa} for recent work in that direction.

That said, a lot of work remains to be done. In our bounce model, we had to use a number of specific functions, which we allowed ourselves to choose freely. An important question is, therefore, how robust are non-singular bounces in a technical sense? That is, within this class of models, what are the minimum requirements for a bounce to occur? Can one formulate a bounce model in supergravity in which the ordinary kinetic term does not have to switch sign (i.e. without a ghost condensate), perhaps based purely on Galileons \cite{Easson:2011zy,Qiu:2011cy,Osipov:2013ssa}? Can one construct models that require less tuning for producing primordial density perturbations with the observed properties? And is it possible to find an explicit embedding of our model in string theory? These are interesting questions that we leave for future work.

\acknowledgements

We would like to thank Justin Khoury, Paul Steinhardt and BingKan Xue for useful discussions. M.K. and J.L.L. gratefully acknowledge the support of the European Research Council via the Starting Grant Nr. 256994 ``StringCosmOS''. B.A.O. is  supported in part by the DOE under contract No. DE-AC02-76-ER-03071and by the NSF under grant No. 1001296. 
%%%%%%%%%%%%%%%%%%%%%%%%%%%%%%%%%%%%%%%%%%%%%%%%%%%%%%%%%%%%%%%%%%%%%%%%%%%%%%%%%%%%%%%%

\appendix
\section{Component Expansion of the Superspace Lagrangian}

In this section, we write out in detail the bosonic component terms of the supergravity Lagrange function
\begin{align}
\mac{L}=-\frac18\int \d^2\Theta2\cE(\bcD^2-8R)\big[&-3\e^{-K(\Phi,\Phid)/3}+\big(\Sib T(\Phi,\Phid,\p_m\Phi,\p_n\Phid,\ldots)\big)\nn\\
&+\big(\cD^\beta\Phi\cDlbP\bcDlb\bcDubPd\Phone(\Phi,\Phid)\big)\big]+H.c.\nn\\
+\int \d^2\Theta2\cE W(\Phi) +H.c.
\label{localcg}
\end{align}
Here $\Phi$ is a chiral superfield with components
\be
\Phi \Blc \equiv A, \qquad \cD^2 \Phi \Blc \equiv - \frac{1}{4}F
\ee
where the bar denotes the lowest component and with $A,F$ being complex scalar fields. The superpotential $W$ is a holomorphic function of $\Phi$ alone, while the K\"{a}hler potential $K$ is a hermitian function of $\Phi$ and $\Phid.$ $T$ is an arbitrary function of $\Phi,\Phid$ and their spacetime derivatives, but with all spacetime indices contracted (it transforms however as a $(2,2)$ tensor in the K\"{a}hler manifold in which the chiral fields take their values -- for details see \cite{Koehn:2012ar}) and the function $G$ appearing in the Galileon term is taken to depend only on $\Phi,\Phid.$
The first summand yields the supergravity version of gravity coupled to a complex scalar field with two-derivative kinetic term and is given, after integration by parts and omission of surface terms, by (cf.~\cite{Wess:1992cp})
\begin{align}
\frac1{e}\mac{L}_X=&\frac1{e}\int \d^2\Theta 2\mac{E}\Big[ \frac{3}{8}(\bcD^2-8R) e^{-K(\Phi,\Phi^\dagger)/3}+W(\Phi)\Big]+H.c. \\
=& e^{-K/3} \big(-\frac12 \cR -\frac13 MM^* +\frac13 b^a b_a\big) \nn\\
& + 3 \Big(\frac{\p^2 e^{-K/3}}{\p A\p A^*}\Big) (\p A \cdot \p A^* - F F^{*}) \nn\\
&  + \I b^m (\p_m A \frac{\p e^{-K/3}}{\p A} - \p_m A^{*} \frac{\p e^{-K/3}}{\p A^{*}}) + MF \frac{\p e^{-K/3}}{\p A}  \\
& + M^* F^{*} \frac{\p e^{-K/3}}{\p A^{*}}-WM^* - W^* M + \p WF + \p W^* F^{*} \nn \ .
\end{align}
The supergravity extension of $X^2T$ is given by
\begin{align}
\mac{L}_{X^2}=&-\frac{1}{8} \int \d^2\Theta 2\mac{E} (\bcD^2-8R)(\Sib T)+H.c.\nn \\
=&8e\big((\p A)^2(\p A^*)^2-2|F|^2(\p A\cdot\p A^*)+|F|^4\big)(T+T^\dagger)\Blc \, .
\end{align}
For the Galileon term, we use the superspace expression
\begin{align}
\frac1{e}\mac{L}_3\blc_\zf=&-\frac{1}{8e}\int \d^2\Theta2\cE(\bcD^2-8R)(\cD^\beta\Phi\cDlbP\bcDlb\bcDubPd\Phone)+H.c.\notag\\
=&\frac1{32}\cDua\cDla\bcDla\bcDua(\cDubP\cDlbP\bcDlb\bcDubPd\Phone)\Blc+\frac1{24}M\cDua\cDla(\cDubP\cDlbP\bcDlb\bcDubPd\Phone)\Blc\nn\\
&+\frac18M^*\bcDla\bcDua(\cDubP\cDlbP\bcDlb\bcDubPd\Phone)\Blc+H.c.\nn\\
=&+8 (\p A)^2\Box A^*\Phone\Blc+8 (\p A^*)^2\Box A\Phone^\dagger\Blc\notag\\
&+\I\frac{16}3(\p A)^2b^mA^*{}_{,m}\Phone\Blc-\I\frac{16}3(\p A^*)^2b^mA_{,m}\Phone^\dagger\Blc\notag\\
&+\I\frac{32}3|F|^2b^mA_{,m}\Phone\Blc-\I\frac{32}3|F|^2b^mA^*{}_{,m}\Phone^\dagger\Blc\notag\\
&+\frac{16}3MF(\p A\cdot\p A^*)\Phone\Blc+\frac{16}3M^*F^*(\p A\cdot\p A^*)\Phone^\dagger\Blc\notag\\
&-\frac{8}3MF|F|^2\Phone\Blc-\frac{{8}}3M^*F^*|F|^2\Phone^\dagger\Blc\notag\\
&+16F^*F^{,m}A_{,m}\Phone\Blc+16FF^{*,m}A^*{}_{,m}\Phone^\dagger\Blc\nn\\
&-2(\p A^*)^2F^*(\cD^2\Phone)\Blc-2(\p A)^2F(\cD^2\Phone)^\dagger\Blc+2|F|^2F(\bcD^2\Phone)\Blc+2|F|^2F^*(\bcD^2\Phone)^\dagger\Blc\nn\\
&-\I4|F|^2A_{,a}\sigma^a_{\a\dot\a}(\cDua\bcDua\Phone)\Blc+\I4|F|^2A^*{}_{,a}\sigma^a_{\a\dot\a}(\cDua\bcDua\Phone)^\dagger\Blc \, ,
\end{align}
where again we omitted all terms including fermions. As it stands, the sum of the above actions is still in Jordan frame due to the coupling of the Ricci scalar to the scalar fields.
To go to Einstein frame, we Weyl rescale $\mac{L}_X+\mac{L}_{X^2}+\mac{L}_3\weyl\mac{L}^W$ according to
\be
{e_n}^a\weyl {e_n}^a\e^{K/6} \ ,
\ee
Under this transformation, the spin connection transforms as
\be
\omega_{nml}\weyl\e^{K/3}(\omega_{nml}+\frac16K_{,m}g_{nl}-\frac16K_{,l}g_{nm}) \ ,
\ee
such that
\begin{align}
\cD_nA_{,b}&\weyl\e^{-K/6}\big(\cD_nA_{,b}-\frac16K_{,n}A_{,b}+\frac16{e_b}^lA^{,m}(K_{,m}g_{nl}-K_{,l}g_{nm})\big)
\end{align}
or simply
\begin{align}
\Box A&\weyl\e^{-K/3}\big(\Box A+\frac13K_{,m}A^{,m}\big).
\end{align}
This leads to the Lagrangian (omitting again total derivatives)
\begin{align}
\frac1{e}\mac{L}^W=&-\frac12\cR-\frac1{12}K^{,m}K_{,m}\nn \\
& + 3\e^{K/3} \Big(\frac{\p^2 \e^{-K/3}}{\p A \p A^{*}}\Big) (\p A\cdot\p A^*) - 3\e^{2K/3}\Big( \frac{\p^2 \e^{-K/3}}{\p A \p A^{*}}\Big) |F|^2\nn \\
& +\frac13 b^m b_m -\frac{\I}3b^m(K_{,A}A_{,m}-K_{,A^*}A^*{}_{,m}) \nn \\
& -\frac13\e^{K/3} MFK_{,A}-\frac13\e^{K/3}M^*F^*K_{,A^*}\nn \\
& -\frac13\e^{K/3} |M|^2 -\e^{2K/3} WM^*-\e^{2K/3}W^* M \nn \\
& +\e^{2K/3}W_{,A} F +\e^{2K/3}W^*_{,A^*} F^{*} \nn \\
& +8\big((\p A)^2(\p A^{*})^2-2\e^{K/3}|F|^2(\p A\cdot\p A^*)+\e^{2K/3} |F|^4\big)(\cT+\cT^*)\notag\\
&+8 (\p A)^2\Box A^*\phone+8 (\p A^*)^2\Box A\phone^*\notag\\
&+\frac83(\p A)^2K^{,m}A^*{}_{,m}\phone+\frac83(\p A^*)^2K^{,m}A_{,m}\phone^*\notag\\
&+\I\frac{16}3(\p A)^2b^mA^*{}_{,m}\phone-\I\frac{16}3(\p A^*)^2b^mA_{,m}\phone^*\notag\\
&+\I\frac{32}3\e^{K/3}|F|^2b^mA_{,m}\phone-\I\frac{32}3\e^{K/3}|F|^2b^mA^*{}_{,m}\phone^*\notag\\
&+\frac{16}3\e^{K/3}MF(\p A\cdot\p A^*)\phone+\frac{16}3\e^{K/3}M^*F^*(\p A\cdot\p A^*)\phone^*\notag\\
&-\frac{8}3\e^{2K/3}MF|F|^2\phone-\frac{{8}}3\e^{2K/3}M^*F^*|F|^2\phone^*\notag\\
&+16\e^{K/3}F^*F^{,m}A_{,m}\phone+16\e^{K/3}FF^{*,m}A^*{}_{,m}\phone^*\nn\\
&-2\e^{K/3}(\p A^*)^2F^*(\cD^2\Phone)^W\Blc-2\e^{K/3}(\p A)^2F(\cD^2\Phone)^{\dagger W}\Blc\nn\\
&+2\e^{2K/3}|F|^2F(\bcD^2\Phone)^W\Blc+2\e^{2K/3}|F|^2F^*(\bcD^2\Phone)^{\dagger W}\Blc\nn\\
&-\I4\e^{K/3}|F|^2A_{,m}\sigma^m_{\a\dot\a}(\cDua\bcDua\Phone)^W\Blc+\I4\e^{K/3}|F|^2A^*{}_{,m}\sigma^m_{\a\dot\a}(\cDua\bcDua\Phone)^{\dagger W}\Blc \, ,
\end{align}
where $\cT\equiv T\Blc{}^W$ and $\phone\equiv G\Blc{}^W$.
In order to disentangle the terms depending on the auxiliary fields $M$ and $F$ we now apply the field redefinition
\be
M=N-K_{,A^{*}}F^{*}+16F^*(\p A\cdot\p A^*)\phone^*-8\e^{K/3}F^*|F|^2\phone^* \, ,
\ee
and obtain
\begin{align}
\frac1{e}\mac{L}^W =&-\frac12 \cR-\frac1{12}K^{,m}K_{,m}\nn \\
& + 3\e^{K/3} \Big(\frac{\p^2 \e^{-K/3}}{\p A \p A^{*}}\Big) (\p A\cdot\p A^*) - 3\e^{2K/3}\Big( \frac{\p^2 \e^{-K/3}}{\p A \p A^{*}}\Big) |F|^2\nn \\
& +\frac13 b^m b_m -\frac{\I}3b^m(K_{,A}A_{,m}-K_{,A^*}A^*{}_{,m}) \nn \\
& + \frac13\e^{K/3} |K_{,A}F|^2 \nn \\
& -\frac13 \e^{K/3} |N|^2 -\e^{2K/3} WN^* - \e^{2K/3}W^* N \nn \\
&+\e^{2K/3} (D_AW)F + \e^{2K/3}(D_AW)^*F^{*}\notag\\
&+8\big((\p A)^2(\p A^*)^2 -2 \e^{K/3} |F|^2 (\p A\cdot\p A^*)+\e^{2K/3} |F|^4\big)(\cT+\cT^*)\notag\\
&-16\e^{2K/3}WF(\p A\cdot\p A^*)\phone-16\e^{2K/3}W^*F^*(\p A\cdot\p A^*)\phone^*\notag\\
&+8\e^KWF|F|^2\phone+8\e^KW^*F^*|F|^2\phone^*\notag\\
&-\frac{16}3\e^{K/3}K_{,A^*}|F|^2(\p A\cdot\p A^*)\phone-\frac{16}3\e^{K/3}K_{,A}|F|^2(\p A\cdot\p A^*)\phone^*\notag\\
&+\frac{8}3\e^{2K/3}K_{,A^*}|F|^4\phone+\frac{8}3\e^{2K/3}K_{,A}|F|^4\phone^*\notag\\
&+\frac{2^8}3\e^{K/3}|F|^2(\p A\cdot\p A^*)^2|\phone|^2-\frac{2^8}3\e^{2K/3}|F|^4(\p A\cdot\p A^*)|\phone|^2+\frac{2^6}3\e^{K}|F|^6|\phone|^2\nn\\
&+8 (\p A)^2\Box A^*\phone+8 (\p A^*)^2\Box A\phone^*\notag\\
&+\frac83(\p A)^2K^{,m}A^*{}_{,m}\phone+\frac83(\p A^*)^2K^{,m}A_{,m}\phone^*\notag\\
&+\I\frac{16}3(\p A)^2b^mA^*{}_{,m}\phone-\I\frac{16}3(\p A^*)^2b^mA_{,m}\phone^*\notag\\
&+\I\frac{32}3\e^{K/3}|F|^2b^mA_{,m}\phone-\I\frac{32}3\e^{K/3}|F|^2b^mA^*{}_{,m}\phone^*\notag\\
&+16\e^{K/3}F^*F^{,m}A_{,m}\phone+16\e^{K/3}FF^{*,m}A^*{}_{,m}\phone^*\nn\\
&-2\e^{K/3}(\p A^*)^2F^*(\cD^2\Phone)^W\Blc-2\e^{K/3}(\p A)^2F(\cD^2\Phone)^{\dagger W}\Blc\nn\\
&+2\e^{2K/3}|F|^2F(\bcD^2\Phone)^W\Blc+2\e^{2K/3}|F|^2F^*(\bcD^2\Phone)^{\dagger W}\Blc\nn\\
&-\I4\e^{K/3}|F|^2A_{,m}\sigma^m_{\a\dot\a}(\cDua\bcDua\Phone)^W\Blc+\I4\e^{K/3}|F|^2A^*{}_{,m}\sigma^m_{\a\dot\a}(\cDua\bcDua\Phone)^{\dagger W}\Blc \ .
\end{align}
The equation of motion for the vector auxiliary field $b_m$ reads
\begin{align}
b_m=&\frac{\I}2(A{}_{,m}K_{,A}-A^{*}{}_{,m}K_{,A^{*}})-16\I\e^{K/3}|F|^2A_{,m}\phone+16\I\e^{K/3}|F|^2A^*_{,m}\phone^*\notag\\
&-8\I (\p A)^2A^*_{,m}\phone+8\I (\p A^*)^2A_{,m}\phone^* \ .
\end{align}
As this equation is algebraic, we are entitled to insert it directly into the Lagrangian, with the result
\begin{align}
\frac1{e}\mac{L}^W =&-\frac12\cR-K_{,AA^*}(\p A\cdot\p A^*)+K_{,AA^*}\e^{K/3}|F|^2\notag\\
&-\frac{16}3\e^{K/3}|F|^2(A^{,m}K_{,A}-A^{*,m}K_{,A^*})(A_{,m}\phone-A^*{}_{,m}\phone^*)\notag\\
&+\frac83(A^{,m}K_{,A}-A^{*,m}K_{,A^*})((\p A^*)^2A_{,m}\phone^*-(\p A)^2A^*{}_{,m}\phone)\notag\\
&-\frac13 \e^{K/3} |N|^2 -\e^{2K/3} WN^* - \e^{2K/3}W^* N \nn \\
&+\e^{2K/3} (D_AW)F + \e^{2K/3}(D_AW)^*F^{*}\notag\\
&+8\big((\p A)^2(\p A^*)^2-2 \e^{K/3} |F|^2 (\p A\cdot\p A^*) + \e^{2K/3} |F|^4\big)(\cT+\cT^*)\notag\\
&-16\e^{2K/3}WF(\p A\cdot\p A^*)\phone-16\e^{2K/3}W^*F^*(\p A\cdot\p A^*)\phone^*\notag\\
&+8\e^KWF|F|^2\phone+8\e^KW^*F^*|F|^2\phone^*\notag\\
&-\frac{16}3\e^{K/3}K_{,A^*}|F|^2(\p A\cdot\p A^*)\phone-\frac{16}3\e^{K/3}K_{,A}|F|^2(\p A\cdot\p A^*)\phone^*\notag\\
&+\frac{8}3\e^{2K/3}K_{,A^*}|F|^4\phone+\frac{8}3\e^{2K/3}K_{,A}|F|^4\phone^*\notag\\
&+8 (\p A)^2\Box A^*\phone+8 (\p A^*)^2\Box A\phone^*\notag\\
&+\frac83(\p A)^2K^{,m}A^*{}_{,m}\phone+\frac83(\p A^*)^2K^{,m}A_{,m}\phone^*\notag\\
&+16\e^{K/3}F^*F^{,m}A_{,m}\phone+16\e^{K/3}FF^{*,m}A^*{}_{,m}\phone^*\nn\\
&-2\e^{K/3}(\p A^*)^2F^*(\cD^2\Phone)^W\Blc-2\e^{K/3}(\p A)^2F(\cD^2\Phone)^{\dagger W}\Blc\nn\\
&+2\e^{2K/3}|F|^2F(\bcD^2\Phone)^W\Blc+2\e^{2K/3}|F|^2F^*(\bcD^2\Phone)^{\dagger W}\Blc\nn\\
&-\I4\e^{K/3}|F|^2A_{,m}\sigma^m_{\a\dot\a}(\cDua\bcDua\Phone)^W\Blc+\I4\e^{K/3}|F|^2A^*{}_{,m}\sigma^m_{\a\dot\a}(\cDua\bcDua\Phone)^{\dagger W}\Blc\nn\\
&+\frac{2^8}3\e^{K/3}|F|^2(\p A\cdot\p A^*)^2|\phone|^2-\frac{2^8}3\e^{2K/3}|F|^4(\p A\cdot\p A^*)|\phone|^2+\frac{2^6}3\e^{K}|F|^6|\phone|^2\nn\\
&+\frac{64}3\big(A^{*,m}(\p A)^2+2\e^{K/3}A^{,m}|F|^2\big)^2\phone^2+\frac{64}3\big(A^{,m}(\p A^*)^2+2\e^{K/3}A^{*,m}|F|^2\big)^2\phone^{*2}\nn\\
&-\frac{128}3\big(A^{*,m}(\p A)^2+2\e^{K/3}A^{,m}|F|^2\big)\big(A_{,m}(\p A^*)^2+2\e^{K/3}A_{*,m}|F|^2\big)|\phone|^2 \, .
\end{align}
The equation of motion for $N$ is also algebraic and remarkably simple, as it is given by
\be
N=-3\e^{K/3}W,
\ee
just as in ordinary two-derivative supergravity. Plugging it back into the Lagrangian, we obtain
\begin{align}
\frac1{e}\mac{L}^W=&-\frac12 \cR -K_{,AA^*}(\p A\cdot\p A^*)+K_{,AA^*}\e^{K/3}|F|^2+3\e^K|W|^2\notag\\
&-\frac{16}3\e^{K/3}|F|^2(A^{,m}K_{,A}-A^{*,m}K_{,A^*})(A_{,m}\phone-A^*{}_{,m}\phone^*)\notag\\
&+\frac83(A^{,m}K_{,A}-A^{*,m}K_{,A^*})((\p A^*)^2A_{,m}\phone^*-(\p A)^2A^*{}_{,m}\phone)\notag\\
&+\e^{2K/3} (D_AW)F + \e^{2K/3}(D_AW)^*F^*\notag\\
&+8\big((\p A)^2(\p A^*)^2-2 \e^{K/3} |F|^2 (\p A\cdot\p A^*) + \e^{2K/3} |F|^4\big)(\cT+\cT^*)\notag\\
&-16\e^{2K/3}WF(\p A\cdot\p A^*)\phone-16\e^{2K/3}W^*F^*(\p A\cdot\p A^*)\phone^*\notag\\
&+8\e^KWF|F|^2\phone+8\e^KW^*F^*|F|^2\phone^*\notag\\
&-\frac{16}3\e^{K/3}K_{,A^*}|F|^2(\p A\cdot\p A^*)\phone-\frac{16}3\e^{K/3}K_{,A}|F|^2(\p A\cdot\p A^*)\phone^*\notag\\
&+\frac{8}3\e^{2K/3}K_{,A^*}|F|^4\phone+\frac{8}3\e^{2K/3}K_{,A}|F|^4\phone^*\notag\\
&+8 (\p A)^2\Box A^*\phone+8 (\p A^*)^2\Box A\phone^*\notag\\
&+\frac83(\p A)^2K^{,m}A^*{}_{,m}\phone+\frac83(\p A^*)^2K^{,m}A_{,m}\phone^*\notag\\
&+16\e^{K/3}F^*F^{,m}A_{,m}\phone+16\e^{K/3}FF^{*,m}A^*{}_{,m}\phone^*\nn\\
&-2\e^{K/3}(\p A^*)^2F^*(\cD^2\Phone)^W\Blc-2\e^{K/3}(\p A)^2F(\cD^2\Phone)^{\dagger W}\Blc\nn\\
&+2\e^{2K/3}|F|^2F(\bcD^2\Phone)^W\Blc+2\e^{2K/3}|F|^2F^*(\bcD^2\Phone)^{\dagger W}\Blc\nn\\
&-\I4\e^{K/3}|F|^2A_{,m}\sigma^m_{\a\dot\a}(\cDua\bcDua\Phone)^W\Blc+\I4\e^{K/3}|F|^2A^*{}_{,m}\sigma^m_{\a\dot\a}(\cDua\bcDua\Phone)^{\dagger W}\Blc\nn\\
&+\frac{2^8}3\e^{K/3}|F|^2(\p A\cdot\p A^*)^2|\phone|^2-\frac{2^8}3\e^{2K/3}|F|^4(\p A\cdot\p A^*)|\phone|^2+\frac{2^6}3\e^{K}|F|^6|\phone|^2\nn\\
&+\frac{2^6}3\big(A^{*,m}(\p A)^2+2\e^{K/3}A^{,m}|F|^2\big)^2\phone^2+\frac{2^6}3\big(A^{,m}(\p A^*)^2+2\e^{K/3}A^{*,m}|F|^2\big)^2\phone^{*2}\nn\\
&-\frac{2^7}3\big(A^{*,m}(\p A)^2+2\e^{K/3}A^{,m}|F|^2\big)\big(A_{,m}(\p A^*)^2+2\e^{K/3}A_{*,m}|F|^2\big)|\phone|^2 \, .
\end{align}
Finally we can derive the equation of motion of the remaining auxiliary field, $F,$ for which we obtain
\begin{align}
0=&K_{,AA^*}F+\e^{K/3}(D_AW)^*+16F\big(\e^{K/3}|F|^2-(\p A\cdot\p A^*)\big)(\cT+\cT^*)\notag\\
&{+16F_{,a}\big(A^{,a}\phone-A^{*,a}\phone^*\big)}\notag\\
&-\frac{16}3F\big(A^{,m}K_{,A}-A^{*,m}K_{,A^*}\big)\big(A_{,m}\phone-A^*{}_{,m}\phone^*\big)-16\e^{K/3}W^*(\p A\cdot\p A^*)\phone^*\notag\\
&+8\e^{2K/3}WF^2\phone+{16}\e^{2K/3}W^*|F|^2\phone^*\notag\\
&-\frac{16}3K_{,A^*}F(\p A\cdot\p A^*)\phone-\frac{16}3K_{,A}F(\p A\cdot\p A^*)\phone^*\notag\\
&+\frac{{32}}3\e^{K/3}K_{,A^*}F|F|^2\phone+\frac{{32}}3\e^{K/3}K_{,A}F|F|^2\phone^*\notag\\
&-\frac{16}3K^{,m}FA^*{}_{,m}\phone^*-16F\Box A^*\phone^*-16FA^*{}_{,m}\phone^{*,m}\nn\\
&-2\e^{K/3}(\p A^*)^2(\cD^2\Phone)^W\Blc-2\e^{K/3}(\p A)^2F\p_{F^*}(\cD^2\Phone)^{\dagger W}\Blc\nn\\
&+2\e^{2K/3}F^2(\bcD^2\Phone)^W\Blc+2\e^{2K/3}|F|^2F\p_{F^*}(\bcD^2\Phone)^W\Blc+4\e^{2K/3}|F|^2(\bcD^2\Phone)^{\dagger W}\Blc\nn\\
&-\I4\e^{K/3}FA_{,m}\sigma^m_{\a\dot\a}(\cDua\bcDua\Phone)^W\Blc+\I4\e^{K/3}FA^*{}_{,m}\sigma^m_{\a\dot\a}(\cDua\bcDua\Phone)^{\dagger W}\Blc\nn\\
&+\frac{2^8}3\e^{K/3}F(\p A\cdot\p A^*)^2|\phone|^2-\frac{2^9}3\e^{2K/3}F|F|^2(\p A\cdot\p A^*)|\phone|^2+2^6\e^{K}F|F|^4|\phone|^2\nn\\
&+\frac{2^9}3\e^{2K/3}F|F|^2[(\p A)^2\phone^2+(\p A^*)^2\phone^{*2}]+\frac{2^8}3\e^{K/3}(\p A\cdot\p A^*)F[(\p A)^2\phone^2+(\p A^*)^2\phone^{*2}]\nn\\
&-\frac{2^{10}}3\e^{2K/3}(\p A\cdot\p A^*)F|F|^2|\phone|^2-\frac{2^9}3\e^{K/3}F(\p A)^2(\p A^*)^2|\phone|^2 \, .
\label{gali-eom}
\end{align}
Here we can see that the equation of motion for $F$ is not algebraic as usual (and moreover it is quintic!), and so it cannot be eliminated directly in general. In our model, we have two cases of interest, and it turns out that in both these cases the equation for $F$ is straightforward to solve: the first is the ekpyrotic phase, where the functions $\cT$ and $\phone$ are zero. In that case, the equation for $F$ can be solved as usual, i.e.
\be
F_{ekpyrotic} = - e^{K/3} K^{,AA^*} (D_A W)^*,
\ee
resulting in the Lagrangian
\begin{align}
\frac1{e}\mac{L}^W_{ekpyrotic}=&-\frac12 \cR -K_{,AA^*}(\p A\cdot\p A^*) - \e^K(K^{,AA^*}|D_AW|^2-3|W|^2).
\end{align}
The second case of interest is the bounce, where the superpotential $W$ is zero. Given that $G$ is a function of $\Phi,\Phid$ only, without derivatives, it follows that $\cD^2 G \Blc \propto F,$ and thus every term in the equation for $F$ contains $F$ itself. This implies that
\be
F_{bounce}=0
\ee
is a valid solution, and is in fact the same solution that would apply in the absence of higher-derivative terms when $W=0.$ Thus this solution corresponds to what in \cite{Koehn:2012ar} we termed the {\it ordinary branch}. Adopting this solution, the Lagrangian during the bounce phase reduces to
\begin{align}
\frac1{e}\mac{L}^W_{bounce}=&-\frac12 \cR -K_{,AA^*}(\p A\cdot\p A^*)+8(\p A)^2(\p A^*)^2(\cT+\cT^*)\notag\\
&+8(\p A)^2\Box A^*\phone+8(\p A^*)^2\Box A\phone^*\notag\\
&+\frac{16}3(\p A)^2(\p A^*)^2\big(K_{,A^*}\phone+K_{,A}\phone^*+4(A^{,m}\phone-A^{*,m}\phone^*)(A_{,m}\phone-A^*{}_{,m}\phone^*)\big) \ .
\end{align}

\bibliographystyle{apsrev}
\bibliography{SuperBounceBib}
\end{document}